\documentclass[preprint,12pt]{elsarticle}
\usepackage{amssymb}
\usepackage{amsmath}
\usepackage{subcaption}
\usepackage{booktabs}
\usepackage{graphicx}
\usepackage{tikz}
\usepackage{multirow}
\usepackage{natbib}

\usepackage{amsthm}

\journal{Nuclear Physics A}

\begin{document}

\begin{frontmatter}

\title{Thermal Evolution of Shape Coexistence in Mo and Ru Isotopes}

\author[a]{Mamta Aggarwal \corref{cor1}}
\ead{mamta.a4@gmail.com}
\author[a]{Pranali Parab}
\author[b]{A. Jain}
\author[c]{G. Saxena}

\address[a]{Department of Physics, University of Mumbai, Vidyanagari, Mumbai - 400098, Maharashtra, India.}
\address[b] {Department of physics, Medicaps University, A.B.Road, Pigdamber, Rau, Indore - 453331, India.}
\address[c]{Department of Physics (H \& S), Govt. Women Engineering College, Ajmer, Rajasthan - 305002, India.}

\cortext[cor1]{Corresponding author}

\begin{abstract}
The temperature-driven shape dynamics of isotopic chains of Mo and Ru elements and their impact on decay modes have been investigated in a statistical theoretical framework with macroscopic-microscopic apporach. These isotopes located at the key points in r-process path are known for the rapid structural changes, shape instabilities and shape coexistence that impact the nuclear processes, decay modes and lifetimes. At high temperatures that may exist in stars or in various nuclear reaction processes, these nuclei undergo a variety of shape and deformation changes due to thermal shell quenching effects influencing the decay energies (Q value), and eventually life-time have been studied in detail. Our findings provide insight into the observed shift in the deformation, shapes and coexisting states due to the diminishing nuclear shell effects in hot nuclei, revealing that the structural changes influence the decay processes and significantly in the astrophysically relevant Mo-Ru region especially around A = 100. 

\end{abstract}

\end{frontmatter}

\section{Introduction}
\label{introduction}

An atomic nucleus, being a finite quantum many-body system, can display a variety of shapes and coexisting shapes~\cite{Heyde11} with competing spherical, axially symmetric prolate and oblate, and triaxial shapes at similar energies~\cite{Bonatsos23}. The intrinsic shape of a nucleus is governed by the interplay of macroscopic bulk properties of nuclear matter and microscopic shell effects which can be profoundly altered by the nuclear temperature which can be as high as up to 2 MeV in stellar environments where nuclear processes generate energy with various nuclear reactions. Since the atomic nuclei undergo a variety of shape transitions under the influence of temperature \cite{MAPRC80, MAPRC90} and affect the dynamics of decay modes and lifetimes, it is therefore important to study the nuclear structural properties of hot nuclei \cite{AgrawalPRC62, MartinPRC68, DudekAPP36} for a wide range of  temperatures \cite{ravlic}. The shape evolution in astrophysically interesting Mo and Ru isotopes has been attracting significant attention in nuclear astrophysics because of the rapidly changing shapes and shape coexistence which impact nuclear stability and decay modes that may further be influenced by the temperature in hot environments. Hence  the study of their structure, masses and decay properties in the ground \cite{MANPA24} and excited states are essential inputs for the astrophysical modeling  \cite{nabi,mishenina} and that needs the attention of nuclear physics and astrophysics researchers in good coordination. \par

Our recent study~\cite{MANPA24} has demonstrated a close correlation between the shape coexistence and nuclear stability with $\beta$-decay lifetime for Mo, Ru isotopic chains in the ground state. With increase in excitation energy, the nuclear energy levels are altered, the single-particle energy states degenerate and the shell effects start to diminish. This leads to shape transition of these hot nuclei from deformed to spherical at a certain temperature known as Critical Temperature $(T_c)$ \cite{MAPRC90} indicating shell quenching. Beyond critical temperature, the nucleon evaporation \cite{ParmarPRC22} may take place followed by $\beta$-decay. To address the influence of finite temperatures on nuclear structural properties and decay modes, we incorporate the temperature degree of freedom to our computations using statistical theory of hot nuclei initiated by Rajasekaran et al. \cite{MTN,MAM} and triaxially deformed Nilson Strutinsky Model with Macroscopic-Microscopic approach. We supress rotational degree of freedom here as the effects of rotation have not been included and we focus only on the temperature effects for this study. Variations in the nuclear structural properties due to the quenching of shell effects that further impact the masses, stability, and level density have been explored. The study involves temperature effects on the structure, shape phase transitions and shape coexisting nuclei  identified in our previous work \cite{MANPA24} in $^{80-124}$Mo and $^{84-126}$Ru isotopic chains with temperature ranging from T = 0.6 MeV - 3 MeV, and the changes in deformation and shape of daughter nuclei and their influence on the possible decay mode, decay energy, level density and lifetimes. \par

\section{Theoretical Formalism}

\label{sec:formalism}
The theoretical framework employed in this work combines the  macroscopic-microscopic aproach of triaxially deformed Nilsson Strutinsky method for ground-state properties with a statistical description for finite-temperature systems.

\subsection{Nilsson Strutinsky Method}

\label{subsec:NSM}

The ground-state properties of a nucleus with proton number $Z$ and neutron number $N$ are obtained using the Nilsson-Strutinsky method. The total energy is expressed as
\begin{equation}
E_{\text{gs}}(Z,N,\beta_2,\gamma) = E_{\text{LDM}}(Z,N) + E_{\text{def}}(Z,N,\beta_2,\gamma) + \delta E_{\text{shell}}(Z,N,\beta_2,\gamma)
\end{equation}
where $E_{\text{LDM}}$ represents the macroscopic liquid-drop model energy, $E_{\text{def}}$ is the deformation energy obtained from surface and coulomb effects, and $\delta E_{\text{shell}}$ denotes the Strutinsky shell correction, all of which are a function of Nilsson's axial quadrupole deformation parameter $\beta_2$ and triaxial parameter $\gamma$.

The macroscopic binding energy $BE_{\text{LDM}}$ is obtained from the Möller–Nix liquid-drop mass formula \cite{PM}, which reproduces experimental binding energies over a wide range of nuclei. The microscopic effects arising due to the non-uniform distribution of nucleons are included through Strutinsky’s shell correction $\delta E_{\text{shell}}$ \cite{VM}, which is defined as
\begin{equation}
\delta E_{\text{shell}} = \sum_{i=1}^{A} \epsilon_i - \tilde{E},
\end{equation}
where $\epsilon_i$ are the single-particle energies obtained from the deformed Nilsson potential, and $\tilde{E}$ represents the smoothed energy obtained by Gaussian averaging with a width parameter of approximately $1.2\hbar\omega$ \cite{MAMI}.

\subsection{Statistical Model of Hot Nuclei}

\label{subsec:finite_temp}
An excited nucleus in thermal equilibrium is treated within a mean-field approximation, where the temperature (T) is taken as input parameter. At finite temperature, the statistical theory of hot nuclei \cite{MNV} is applied through the grand canonical partition function for a deformed nucleus given by
\begin{equation}
Q(\alpha_Z,\alpha_N,\beta') = \sum_i \exp(-\beta' E_i + \alpha_Z Z_i + \alpha_N N_i),
\end{equation}
where $\beta' = 1/T$ (with $k_B = 1$), and $\alpha_Z$, $\alpha_N$ are Lagrange multipliers ensuring the conservation of energy and the average proton and neutron numbers respectively. The thermodynamic expectation values are obtained from the saddle-point equations:
\begin{align}
\frac{\partial \ln Q}{\partial \beta'} &= \langle E \rangle \\
\frac{\partial \ln Q}{\partial \alpha_Z} &= \langle Z \rangle \\
\frac{\partial \ln Q}{\partial \alpha_N} &= \langle N \rangle.
\end{align}

The single-particle energy eigenvalues $\epsilon_i^Z$ and $\epsilon_i^N$ \cite{HM} of the deformed Nilsson Hamiltonian, diagonalized in a cylindrical basis \cite{GS,EIS} are taken as input for the statistical calculations. The corresponding thermodynamic expectation values are given by
\begin{align}
\langle Z \rangle &= \sum_i n_i^Z = \sum_i [1 + \exp(-\alpha_Z + \beta' \epsilon_i^Z)]^{-1}\\
\langle N \rangle &= \sum_i n_i^N = \sum_i [1 + \exp(-\alpha_N + \beta' \epsilon_i^N)]^{-1}\\
\langle E(T) \rangle &= \sum_i n_i^Z \epsilon_i^Z + \sum_i n_i^N \epsilon_i^N.
\end{align}

where $n_i$ is the occupation probability. The excitation energy of the nucleus is calculated as
\begin{equation}
E^*(T) = E(T) - E(0),
\end{equation}
where $E(0)$ is the ground-state energy,
\begin{equation}
E(0) = \sum \epsilon_i^Z + \sum \epsilon_i^N.
\end{equation}

Entropy and free energy of the system are derived from the statistical properties of the single-particle levels at finite temperature. The entropy is calculated as
\begin{equation}
S = -\sum_i [n_i \ln n_i + (1 - n_i) \ln(1 - n_i)]. 
\end{equation}
We define an effective excitation energy $U_{eff}(T)$ as
\begin{equation}
U_{eff}(T) = E^*(T) -\delta E_{shell},
%\label{equation}
\end{equation}
where part of the excitation energy is used to overcome the shell forces 
which are deformation dependant. The quantity $\delta$E$_{shell}$ is the ground state shell correction obtained using Strutinsky's prescription ~\cite{VM}. The level density parameter 'a' is computed using effective excitation energy $U_{eff}(T)$ as
\begin{equation}
 a(T) = U_{eff}(T)/T^2
%\label{equation}
\end{equation}

\subsection{Equilibrium Shapes at Finite Temperature}

\label{subsec:equilibrium}

The shape is parameterized by the axial deformation parameter $\beta_2$ which ranges from 0.0 to 0.4, while the angular  deformation parameter $\gamma$ varies from $-180^\circ$ (oblate non-collective) to $-120^\circ$ (prolate collective), $-60^\circ$ (oblate collective) to $0^\circ$ (prolate non-collective), with triaxial configurations lying in between. Minimization of the free energy $F$ with respect to $\beta_2$ and $\gamma$ \cite{MAI} yields the most probable equilibrium configuration of the nucleus at a given temperature.
\begin{equation}
F(Z,N,\beta_2,\gamma,T) = E(Z,N,\beta_2,\gamma,T) - T*S(Z,N,\beta_2,\gamma,T).
\end{equation}

The total energy at finite temperature is obtained by combining the ground-state energy ($E_{\text{gs}} = -BE_{\text{gs}}$) from the Nilsson Strutinsky Method with the thermal excitation energy derived from the statistical model as
\begin{align}
E(Z,N,\beta_2,\gamma,T) &= -BE(Z,N,\beta_2,\gamma,T) \nonumber \\
&= -BE_{\text{gs}}(Z,N) + E^*(T,\beta_2,\gamma) \nonumber \\
&= -BE_{\text{LDM}}(Z,N) + E_{\text{def}}(Z,N,\beta_2,\gamma) \nonumber \\
& + \delta E_{\text{shell}}(Z,N,\beta_2,\gamma) + E^*(Z,N,\beta_2,\gamma,T).
\end{align}

\section{Result and Discussion}
The evolution of nuclear shape with temperature offers insights into the interplay between shell structure and equilibrium deformation. The quadrupole moments of the excited states and electromagnetic transition rates  between them are observables used to study nuclear shapes experimentally and to verify theoretical nuclear structure model predictions \cite{Wood92}. 

\begin{figure}[htbp]
 	\begin{subfigure}[t]{0.535\linewidth}
 		\centering
 		\includegraphics[width=\linewidth]{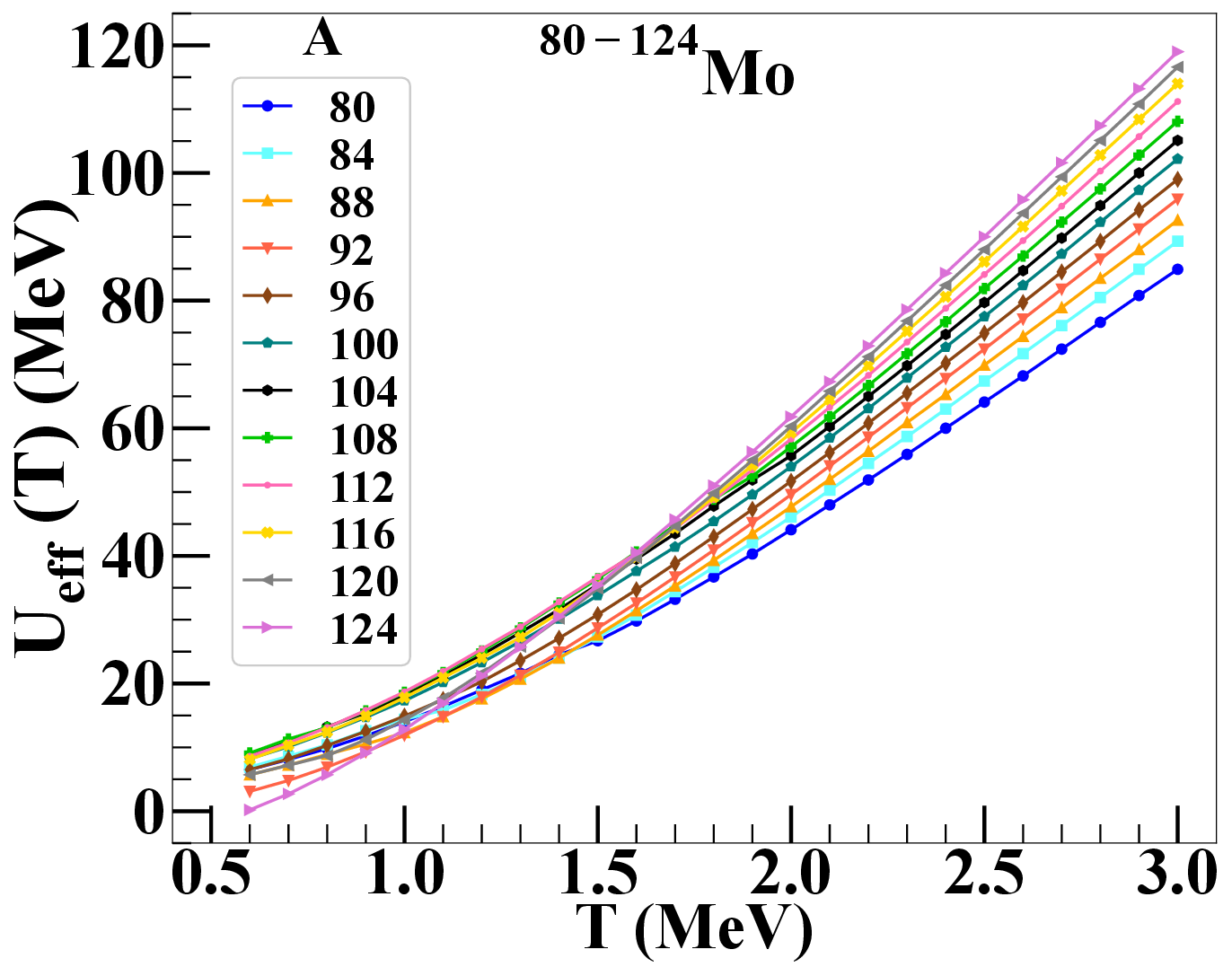}
 		 \caption{Mo Isotopes}
 		 \label{ExMo}
 	\end{subfigure}
 	\begin{subfigure}[t]{0.465\linewidth}
 		\centering
 		\includegraphics[width=\linewidth]{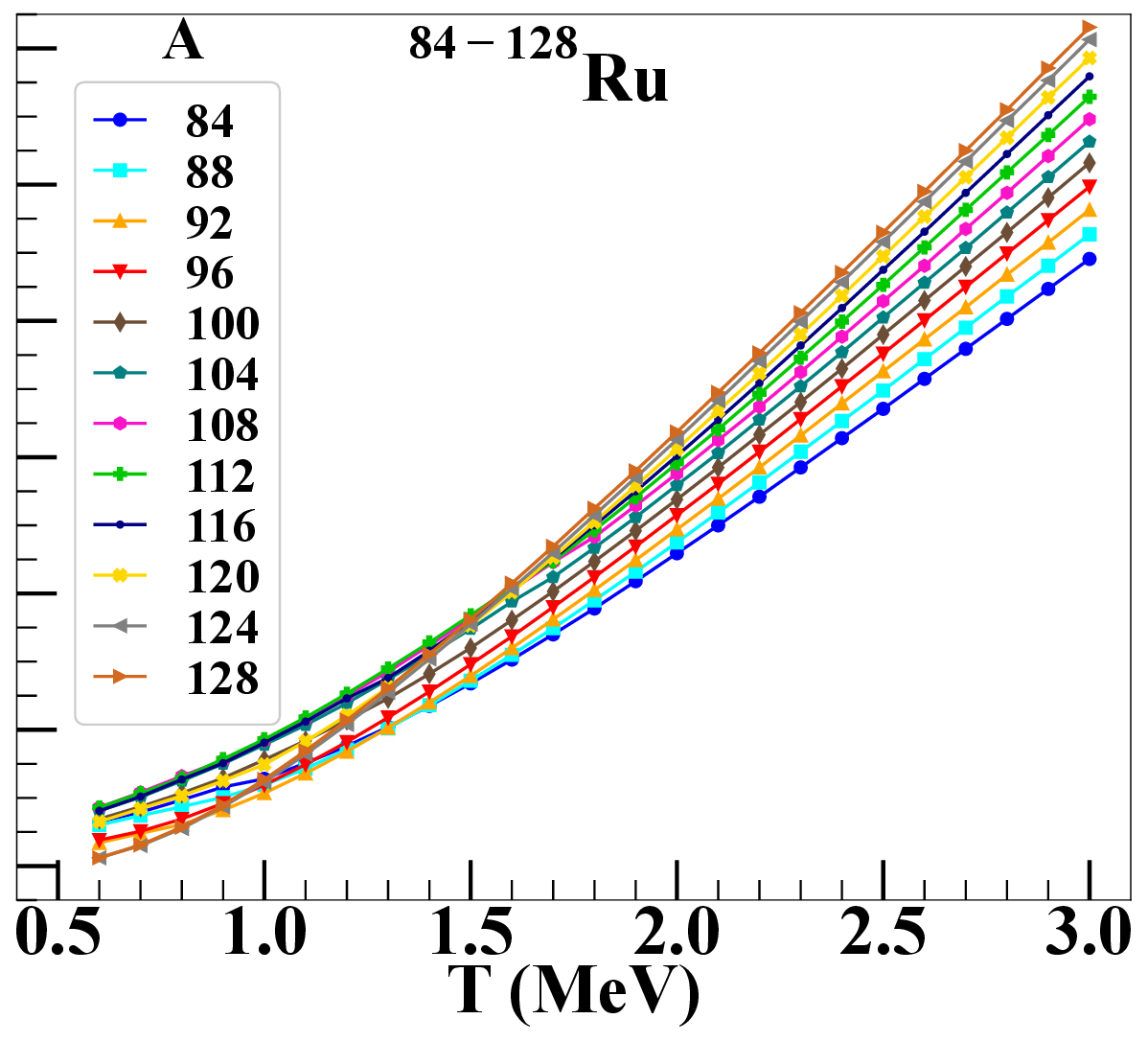}
 		\caption{Ru Isotopes}
 		 \label{ExRu}
 	\end{subfigure}
 	\caption{Temperature corresponding to Excitation energy.}
 	\label{ExT}
 \end{figure}

Effective excitation energy $U_{eff}(T)$ corresponding to temperature (T), input parameter for our calculation, is shown in Fig.~\ref{ExT}, which demonstartes that the excitation energy $U_{eff}(T)$ for Mo and Ru isotopes increases smoothly with temperature. \par

 \begin{figure}[htbp]
 	\begin{subfigure}[t]{0.5\linewidth}
 		\centering
 		\includegraphics[width=\linewidth]{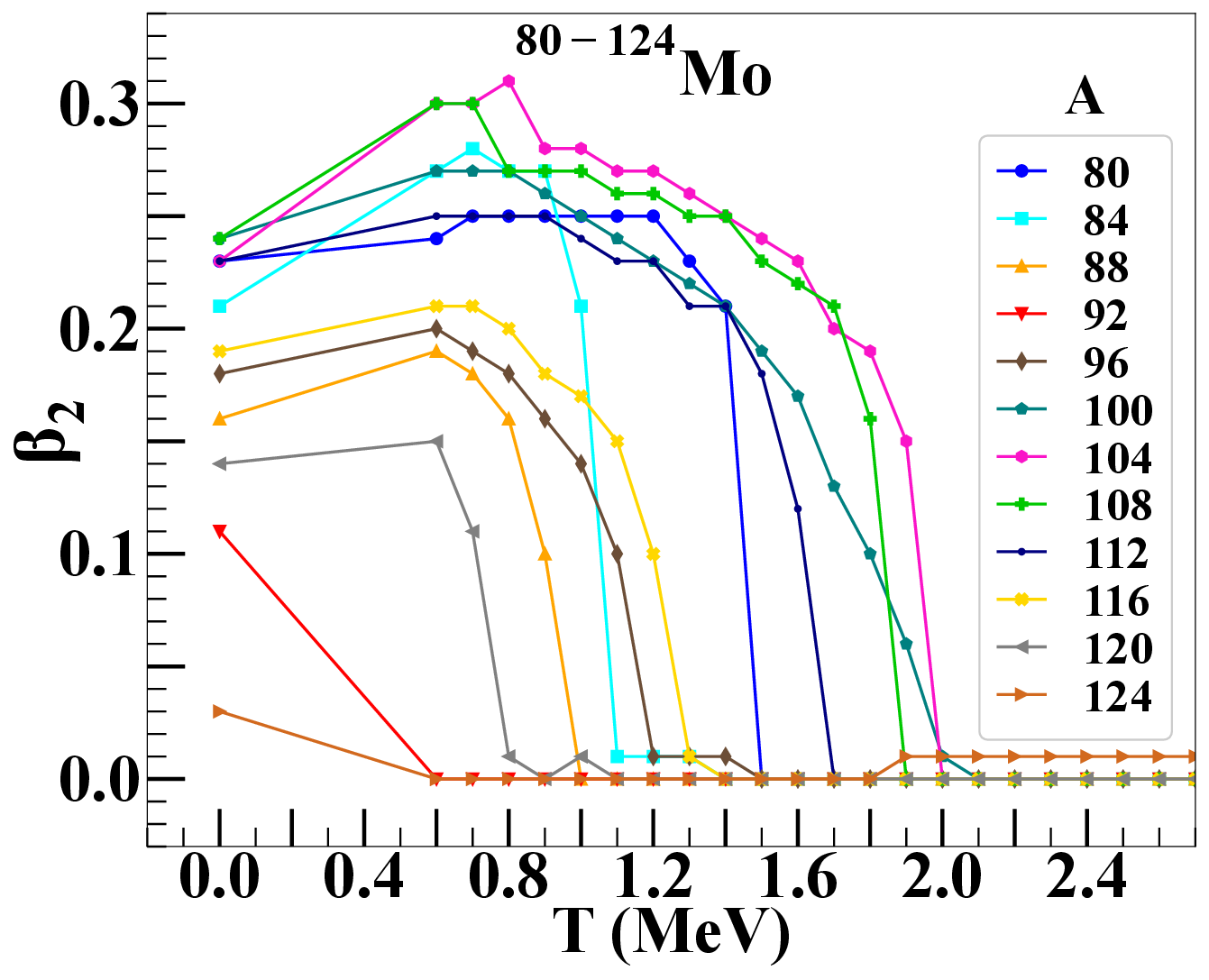}
 		 \caption{Mo Isotopes}
 		 \label{tcMo}
 	\end{subfigure}
 	\begin{subfigure}[t]{0.5\linewidth}
 		\centering
 		\includegraphics[width=\linewidth]{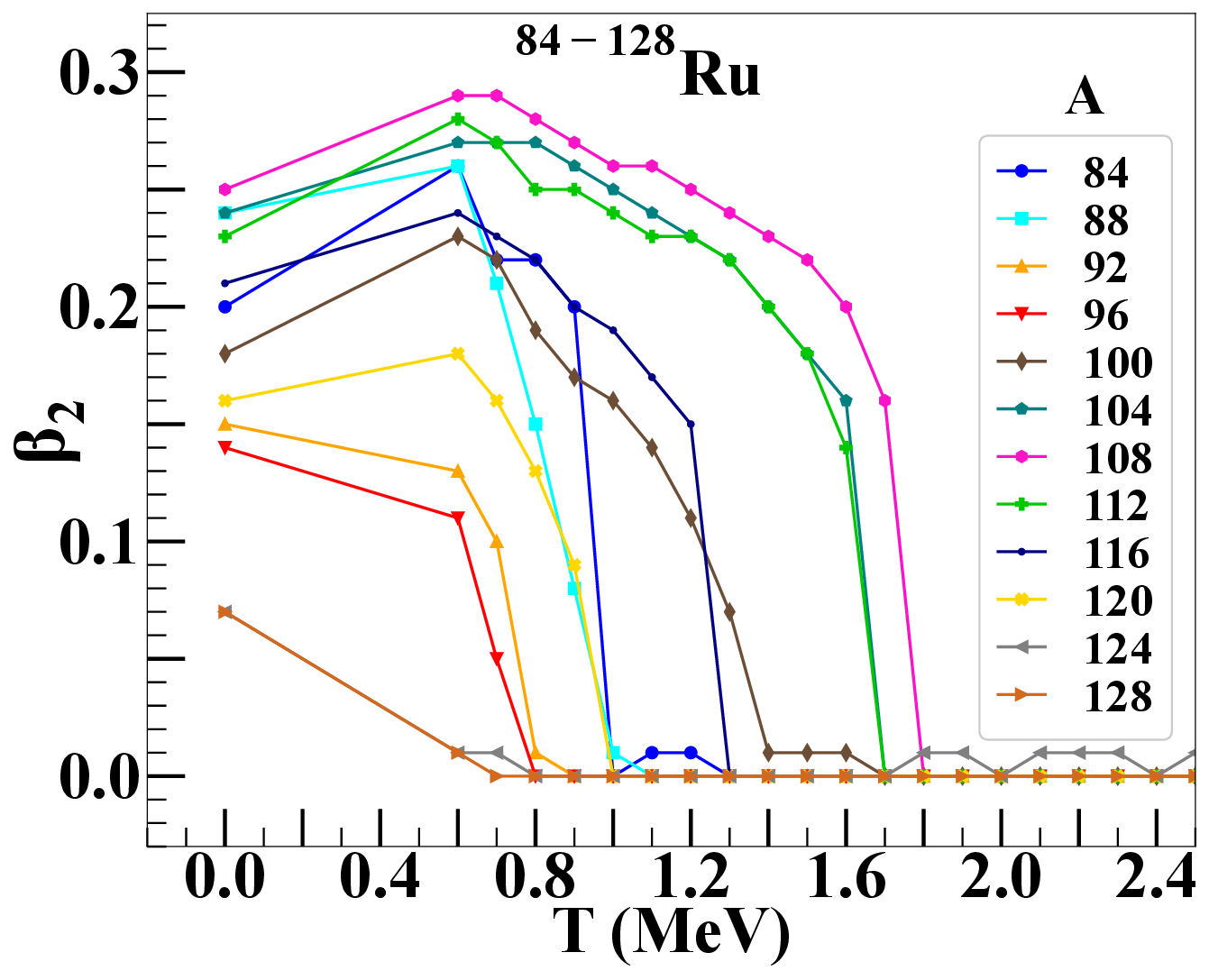}
 		\caption{Ru Isotopes}
 		 \label{tcRu}
 	\end{subfigure}
 	\caption{Variation of deformation parameter $\beta_2$ with Temperature.}
 	\label{Tc}
 \end{figure}

The variation of equilibrium deformation $\beta_2$  with temperature for the isotopic chain of Mo and Ru in mass region  80-120 is displayed in figure~\ref{Tc}. As seen in the figure, nuclei exhibit finite deformation at lower temperatures, with mostly oblate or triaxial shapes and approach sphericity while approaching magic nuclei with N $=$ 50, 82 shell closures. As temperature rises, characteristic sharp reduction in the quadrupole deformation parameter $\beta_2$ is observed. At a particular temperature known as Critical Temperature ($T_c$),  $\beta_2$ values tend to zero and the nucleus transitions to a nearly spherical shape. The observed value of $T_c$ = 2.0 MeV for $^{104}\mathrm{Mo}$, a maximum value of $T_c$ around mid shell nuclei, start reducing while approaching the magic nuclei, decreasing to a value of 0.7 MeV at shell closures N $=$ 50 and 82. Closed shell nucleus $^{92}\mathrm{Mo}$ is observed to be spherical at a temperature of 0.6 MeV. Region around near N = 60 mid-shell nuclei where several nuclei exhibited shape coexistence in ground state (T $=$ 0 MeV) \cite{MANPA24} with large deformation, the critical temperature shows a very high value, in fact one of the highest among Mo isotopes. Around shell closures, the deformation is very low, hence critical temperature is low and the nucleus attains sphericity at a much lower excitation. Critical temperatures for Ru isotopes deduced from Fig \ref{tcRu} show $T_c$ = 1.8 MeV for $^{108}\mathrm{Ru}$, a maximum value around mid-shell nucleus at N = 60 as observed in Mo isotopes as well. Afterwards it starts decreasing and assumes a value 0.7 MeV, lowest for the isotopic chain around neutron shell closure N $=$ 50, 82 indicating that the critical temperatures are different for nuclei with different structural dynamics and get affected by shell effects.\par 

As temperature starts increasing (as seen in Fig.~\ref{Tc}), the deformation rises initially that may be due to the particle rearrangement near Fermi level at low temperatures with more particles going to higher levels. With further increase in T, the shell quenching effects starts dominating and the equilibrium shape starts decreasing and eventually drops to zero at critical temperature $T_c$.\par

The phenomenon of shape coexistence, a notable structural feature prevalent in this part of nuclear chart, has become a useful paradigm to explain the competition between the monopole part of the nuclear effective force that tends to stabilize the nucleus into a spherical shape around shell closures, and the strong correlations that favor the nucleus into a deformed shapes around mid-shell regions. This interplay manifests in the low-lying structure of Mo and Ru isotopes that exhibit a evolution with temperature, reflecting underlying changes in nuclear deformation.\par 

At low excitation energies, both Mo and Ru nuclei display pronounced quadrupole collectivity, characterized by the coexistence of spherical, axially symmetric, and triaxial configurations. Self-consistent DFT calculations with the UNEDF0 functional predict triaxial ground-state deformations for Mo and Ru isotopes with  A = 106 to 112 \cite{ZhangPRC92}. Few experimental studies indicate a triaxial ground state and an excited prolate state for Mo isotopes around A = 100 \cite{ZielinskNPA712,WrzosekPRC86}. Measurements of low-lying collective bands and E2/M1 transition strengths \cite{EldrigeEPJA54} further confirm the quadrupole nature of these excitations. Overall, the observed trends are broadly consistent with our mean-field results, where we found shape transitions and coexistence between axially symmetric and triaxial shapes in ground state \cite{MANPA24} and excited states as well.

\begin{figure}[htbp]
  \centering
  \begin{tikzpicture}
    \def\w{3.3cm}  % width of each small image
    \def\h{3cm}  % height of each small image

    % row 1
    \foreach \c [count=\k from 0] in {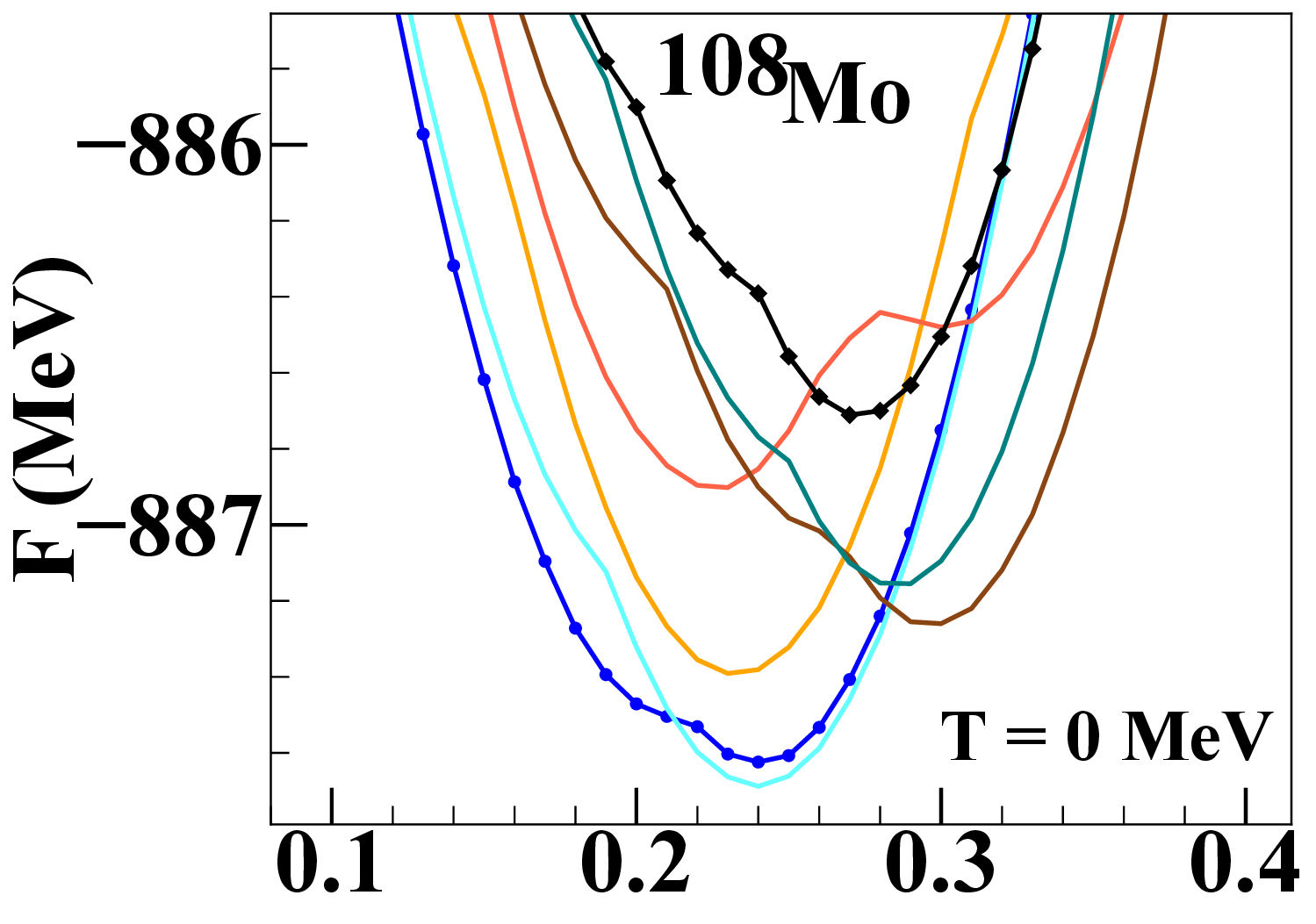,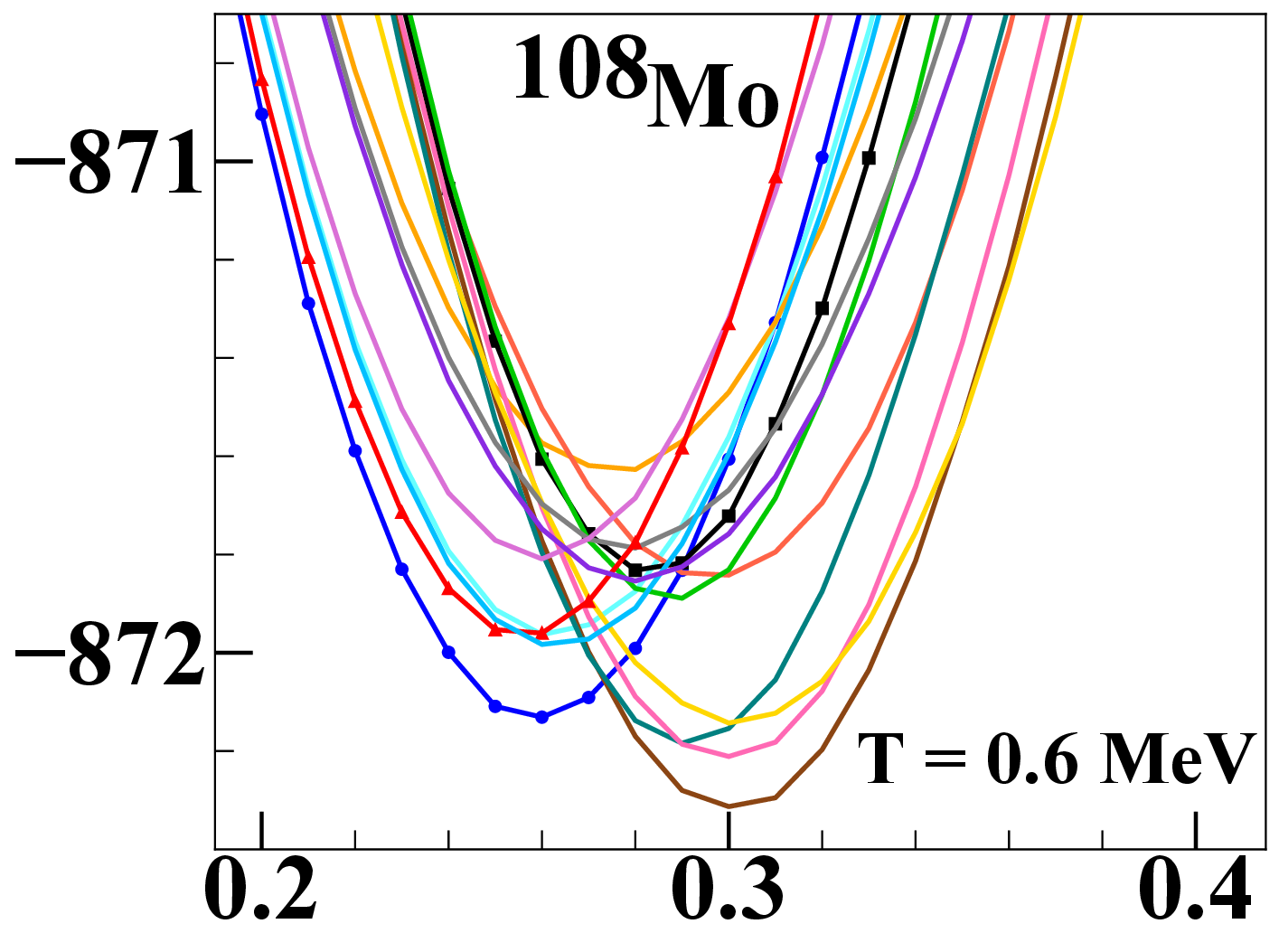, 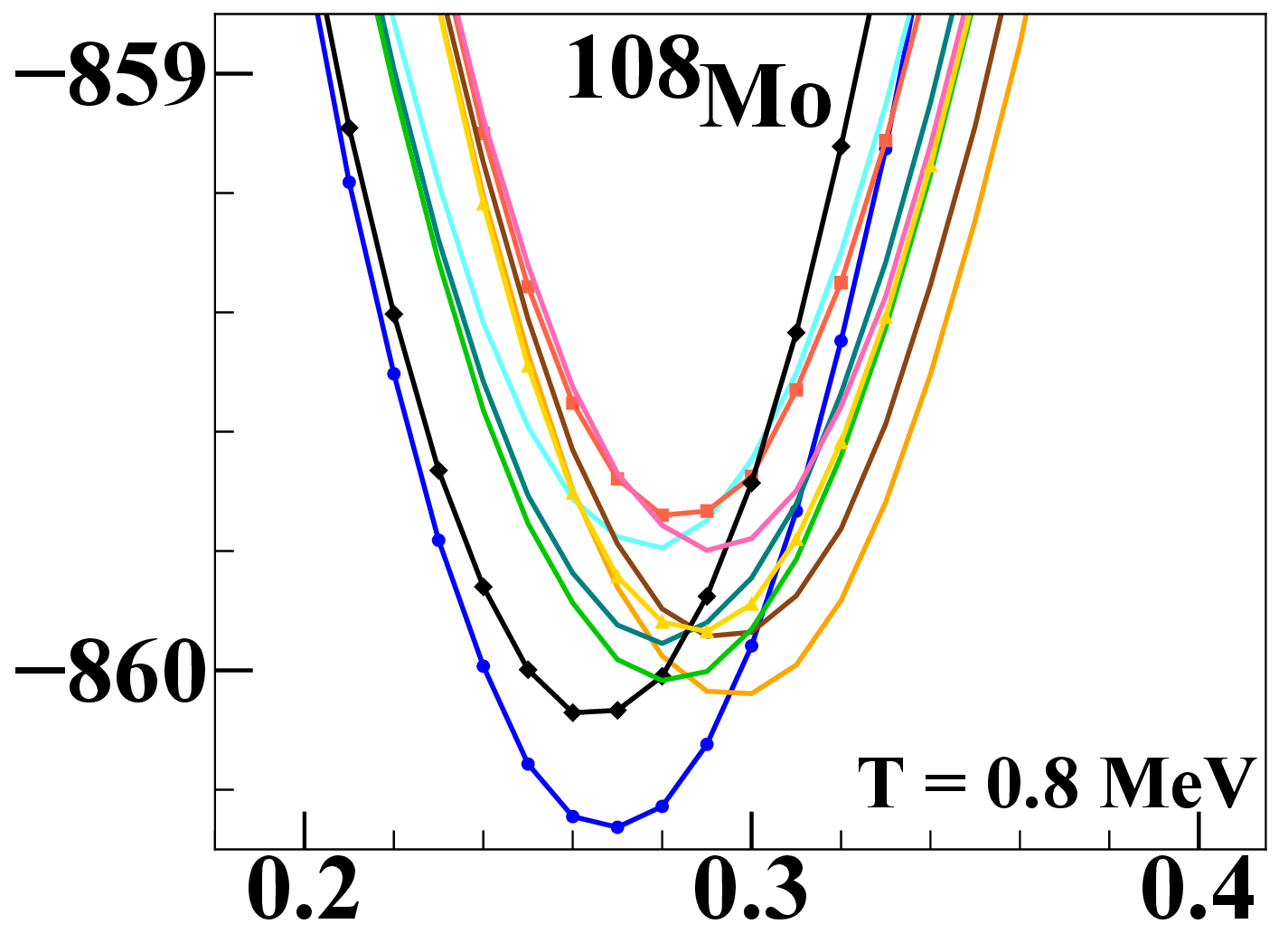,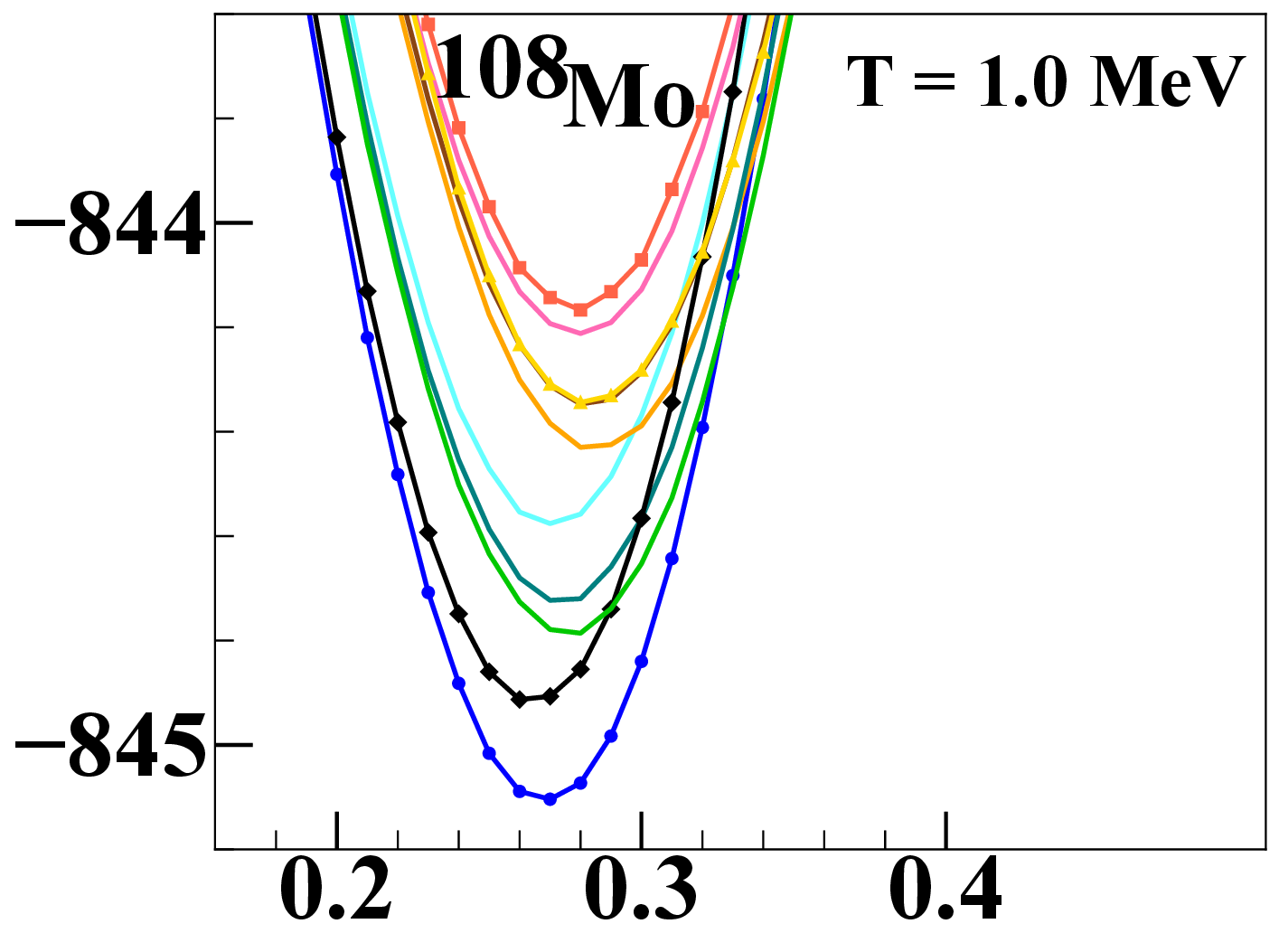}
      {
        \node[anchor=north west, inner sep=0] (s\k) at ({\k*\w},0)
          {\includegraphics[width=\w,height=\h]{\c}};
      }%

    % row 2
    \foreach \c [count=\k from 0] in {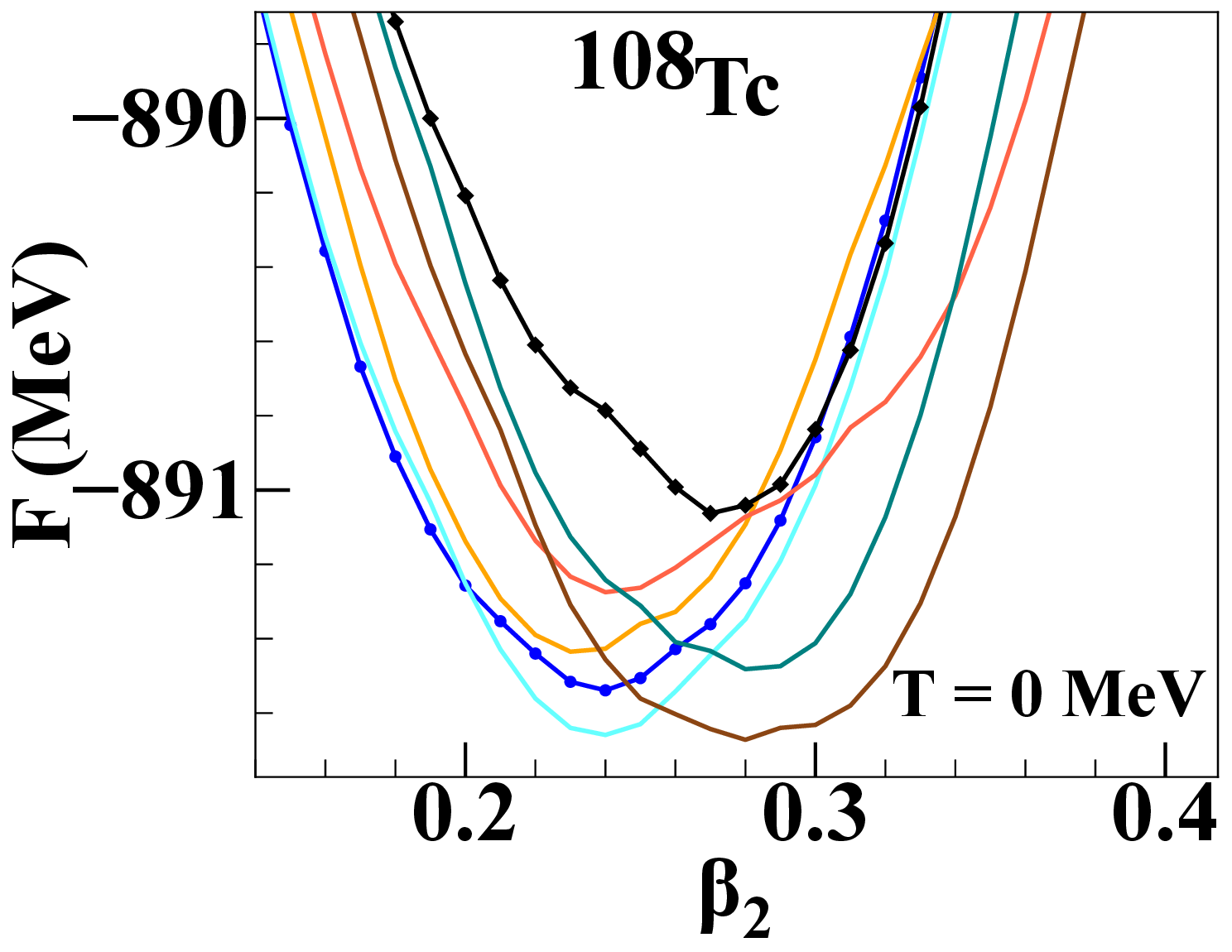,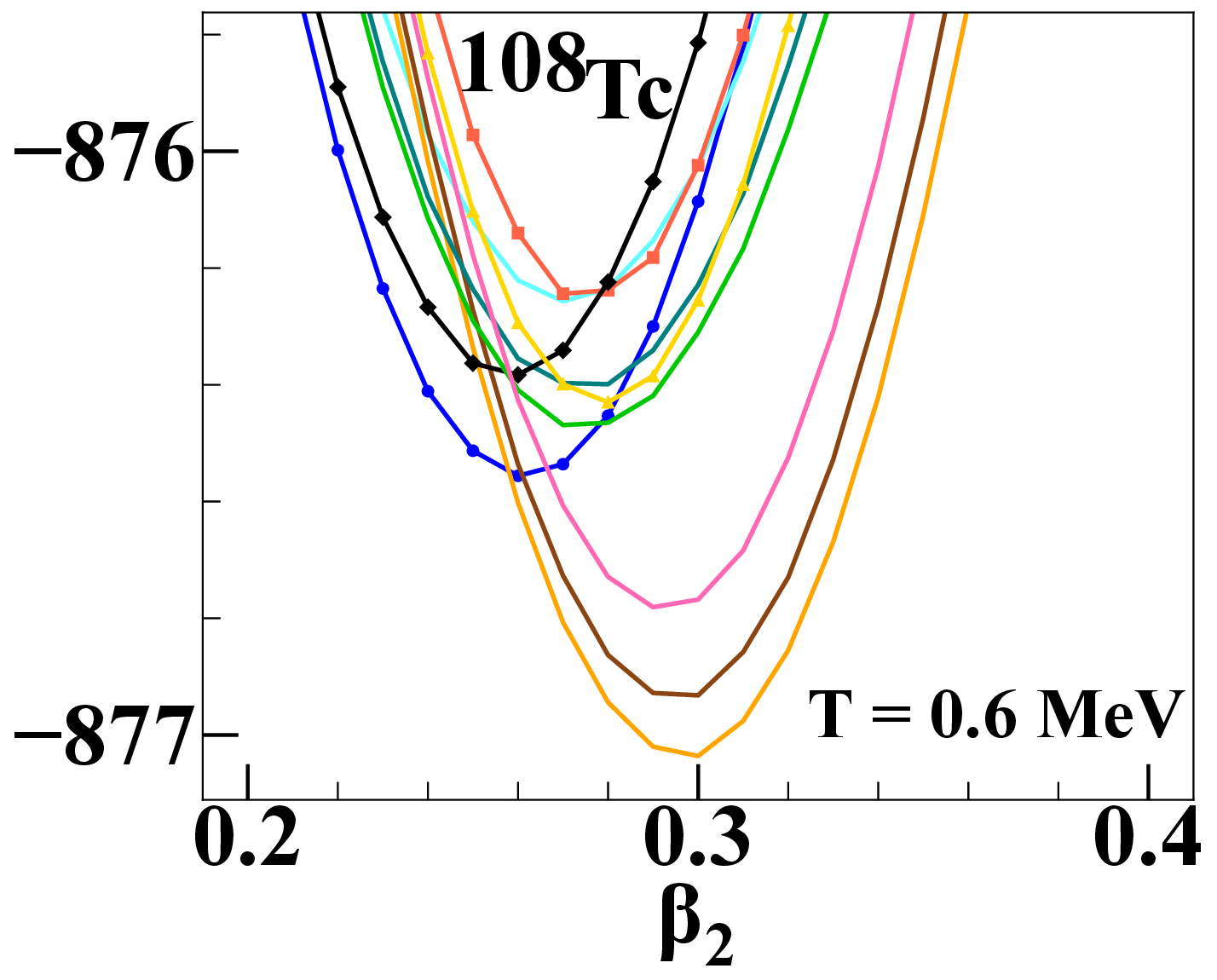,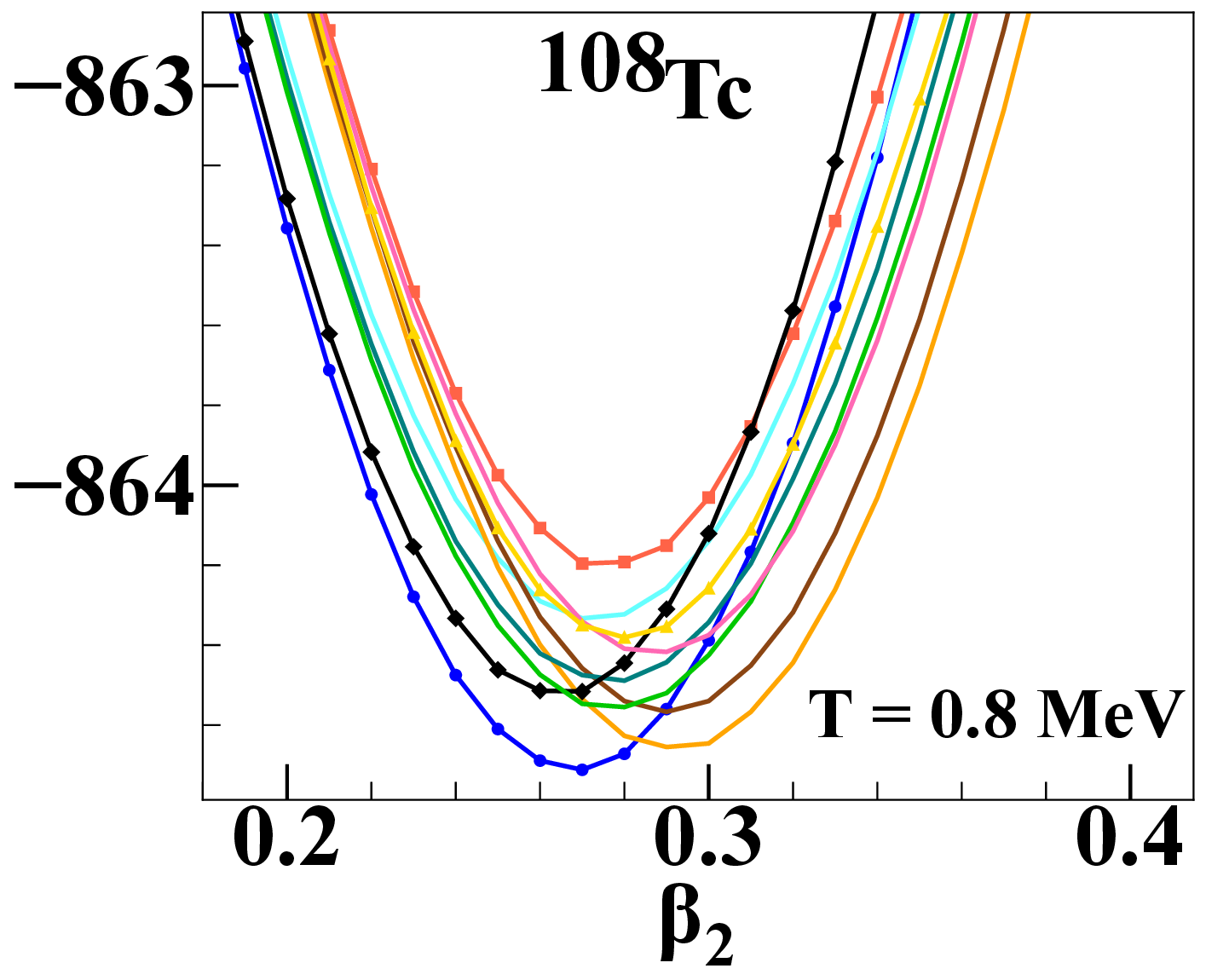,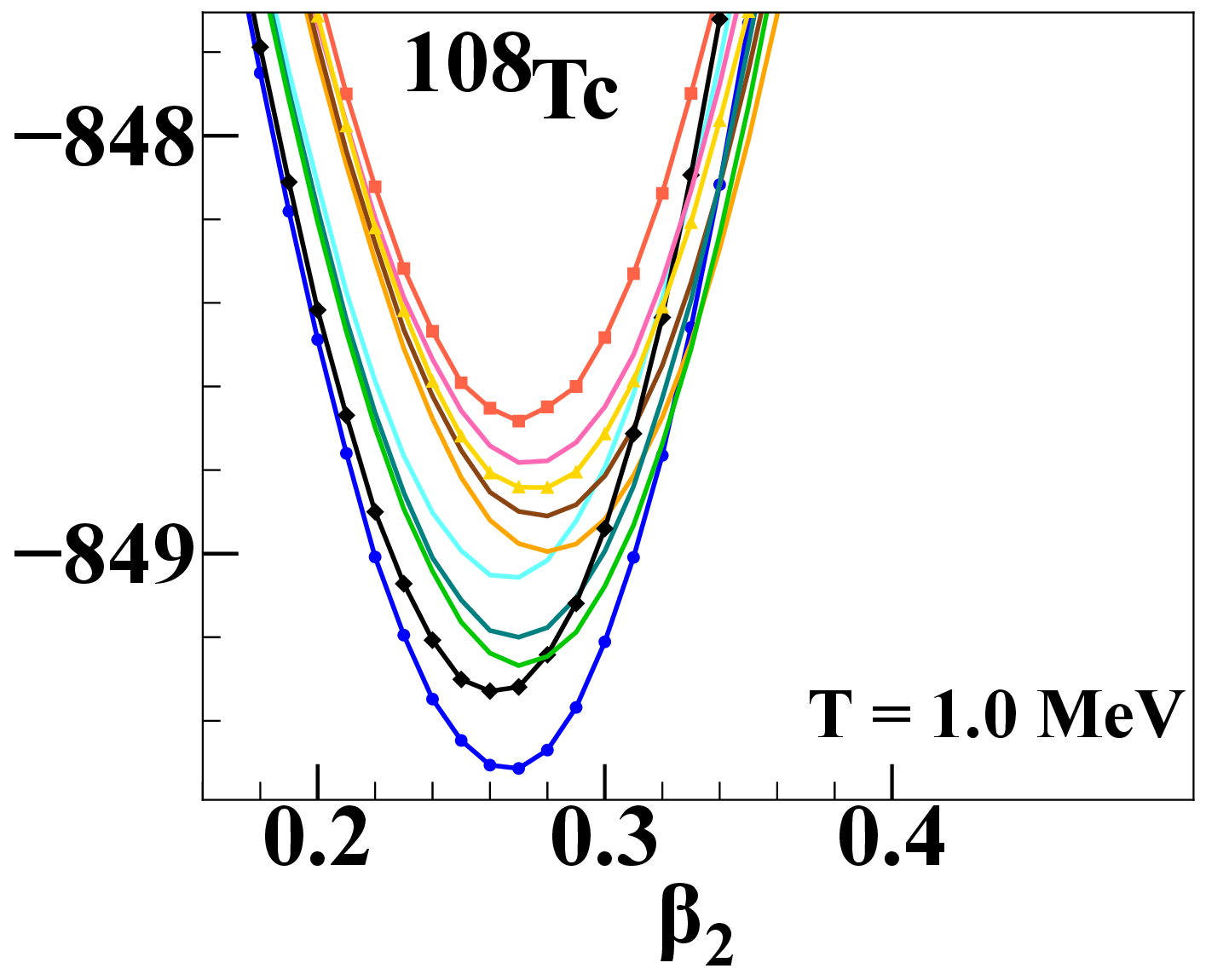}     
     {
        \node[anchor=north west, inner sep=0] (s2\k) at ({\k*\w},-\h-0.2cm)
          {\includegraphics[width=\w,height=\h]{\c}};
      }

    % tall image at right 
%  
\node[anchor=north east, inner sep=0] (tall) at ({3.53*\w + 1.5 cm}, -0.7)
  {\includegraphics[height=4.2cm,keepaspectratio]{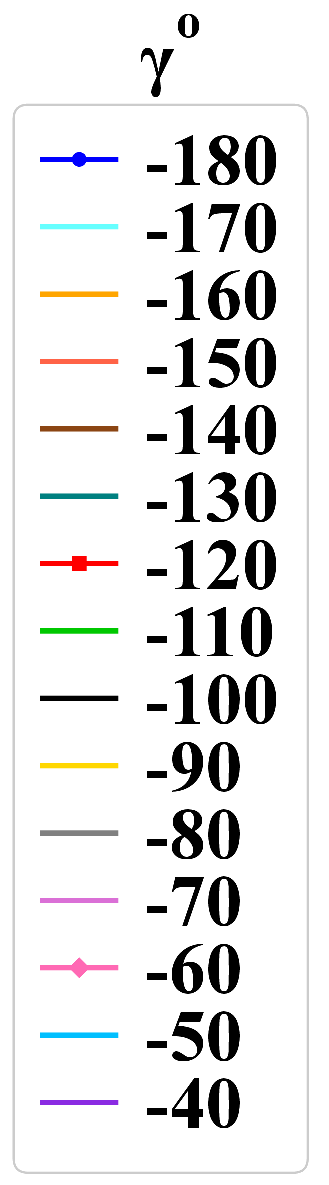}};

  \end{tikzpicture}

  \caption{Shape evolution of $^{108}\mathrm{Mo}$ and its $\beta^-$decay daughter $^{108}\mathrm{Tc}$}
  \label{td108MoTc}
\end{figure}

\begin{figure}[htbp]
  \centering
  \begin{tikzpicture}
    \def\w{3.3cm}  % width of each small image
    \def\h{3cm}  % height of each small image

    % row 1
    \foreach \c [count=\k from 0] in {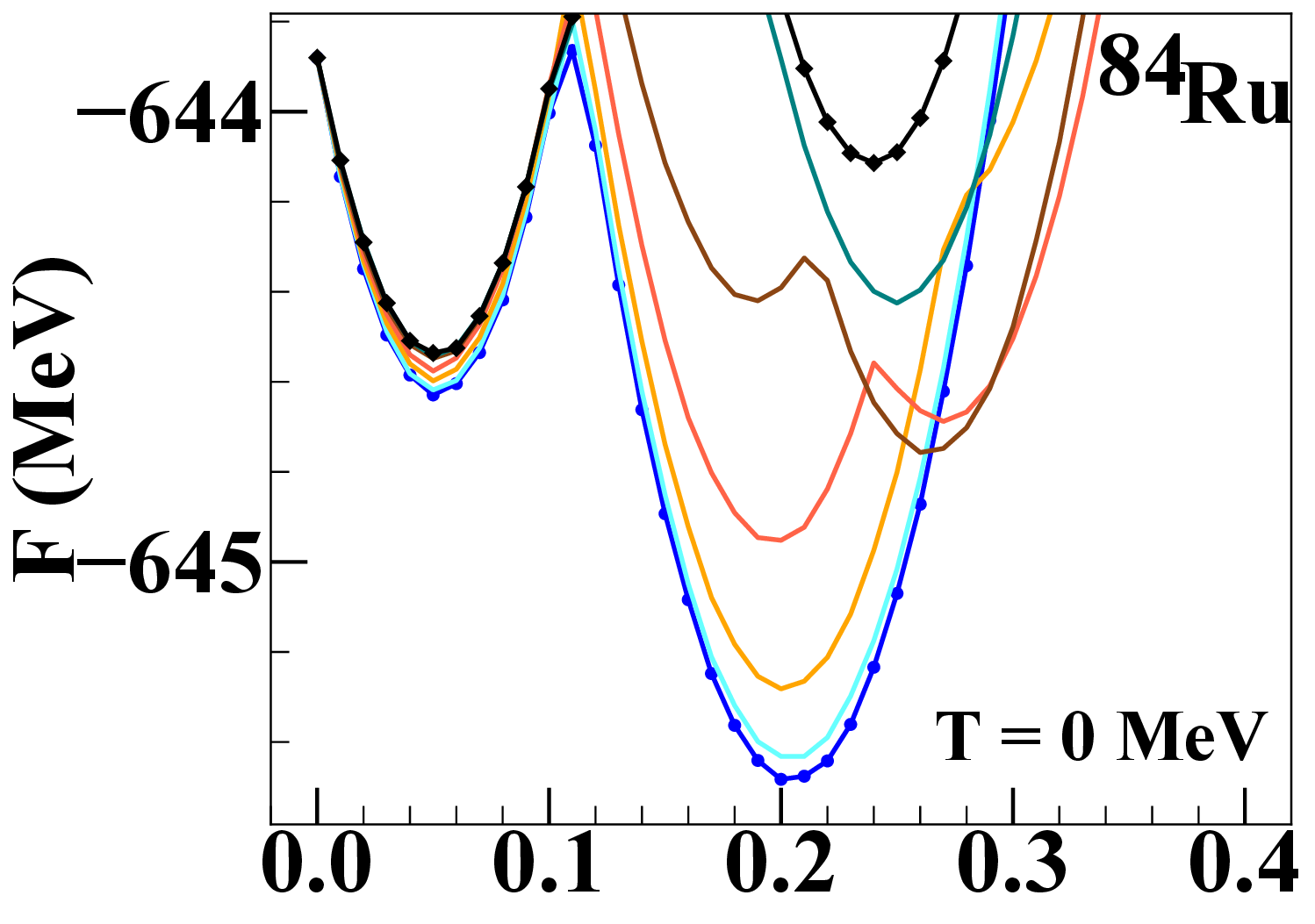, 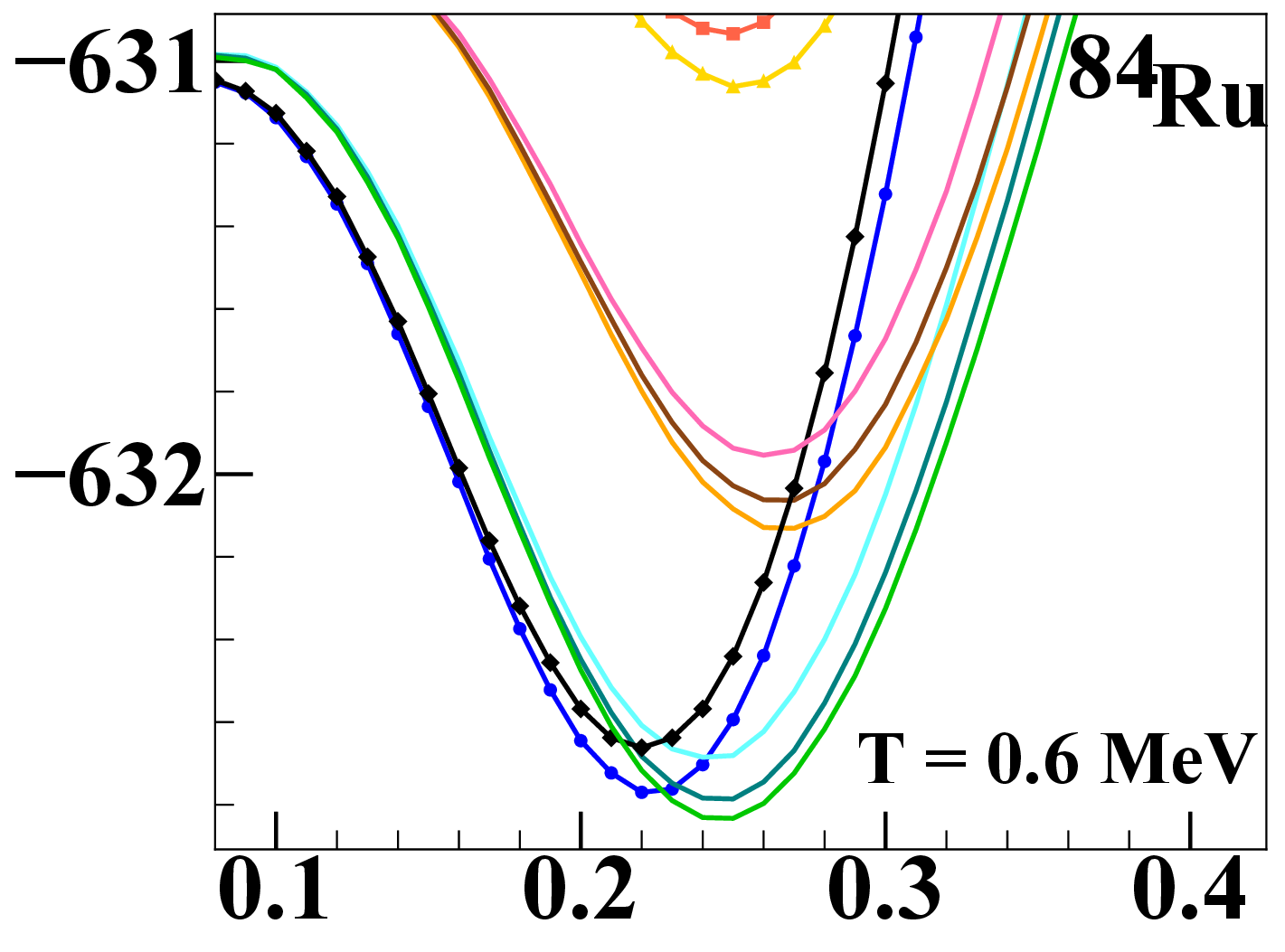, 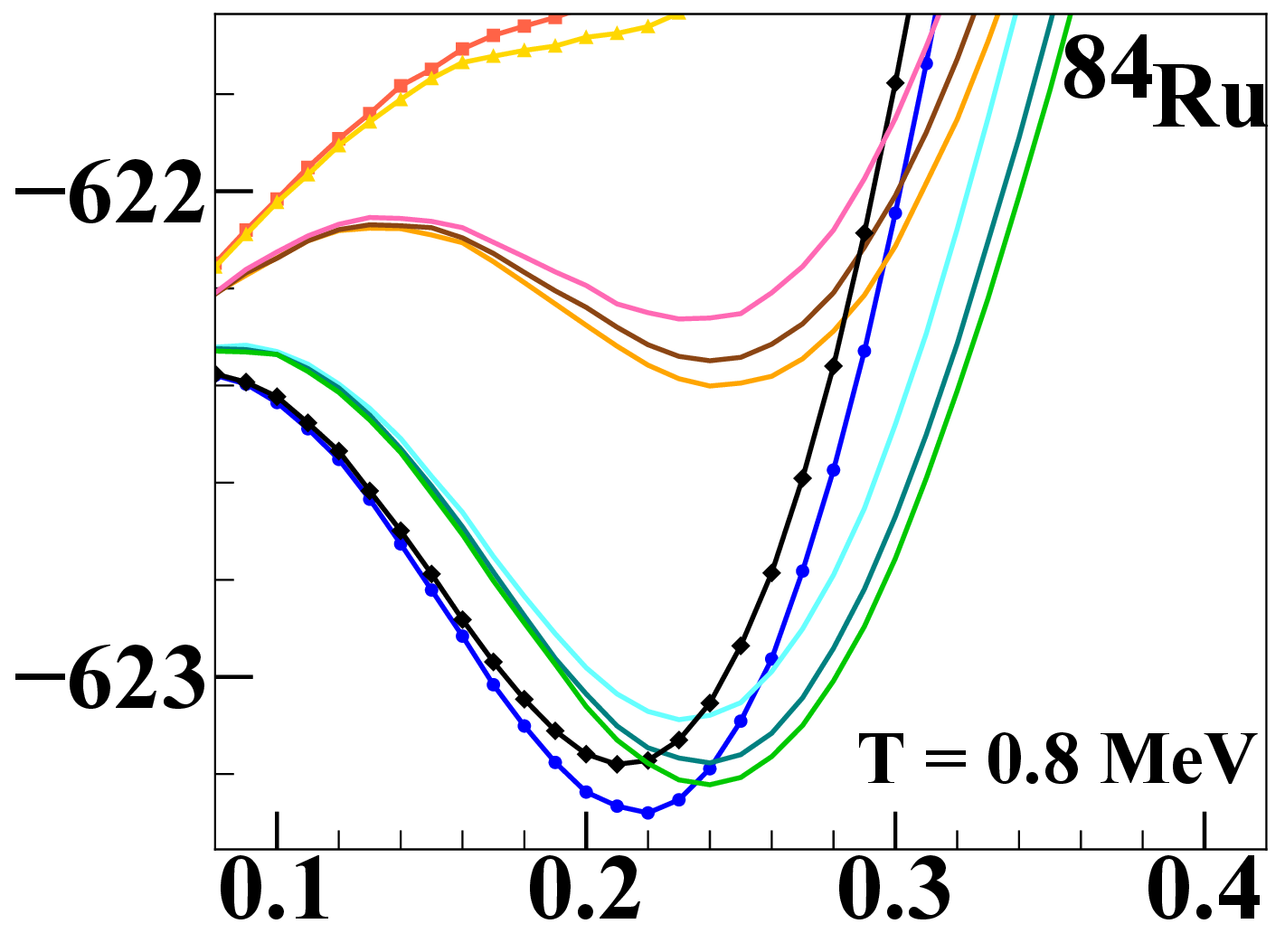,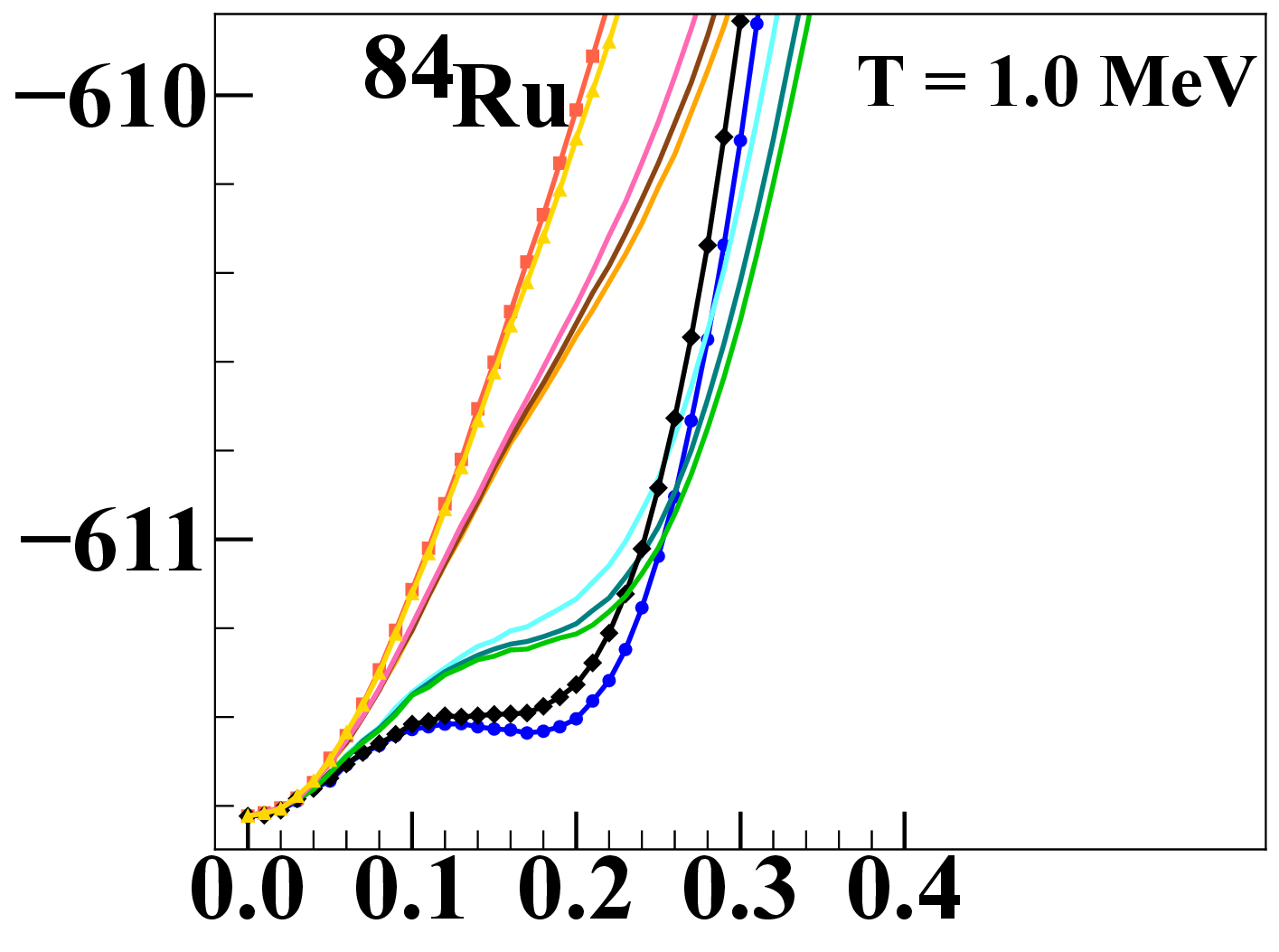}
      {
        \node[anchor=north west, inner sep=0] (s\k) at ({\k*\w},0)
          {\includegraphics[width=\w,height=\h]{\c}};
      }

    % row 2
    \foreach \c [count=\k from 0] in {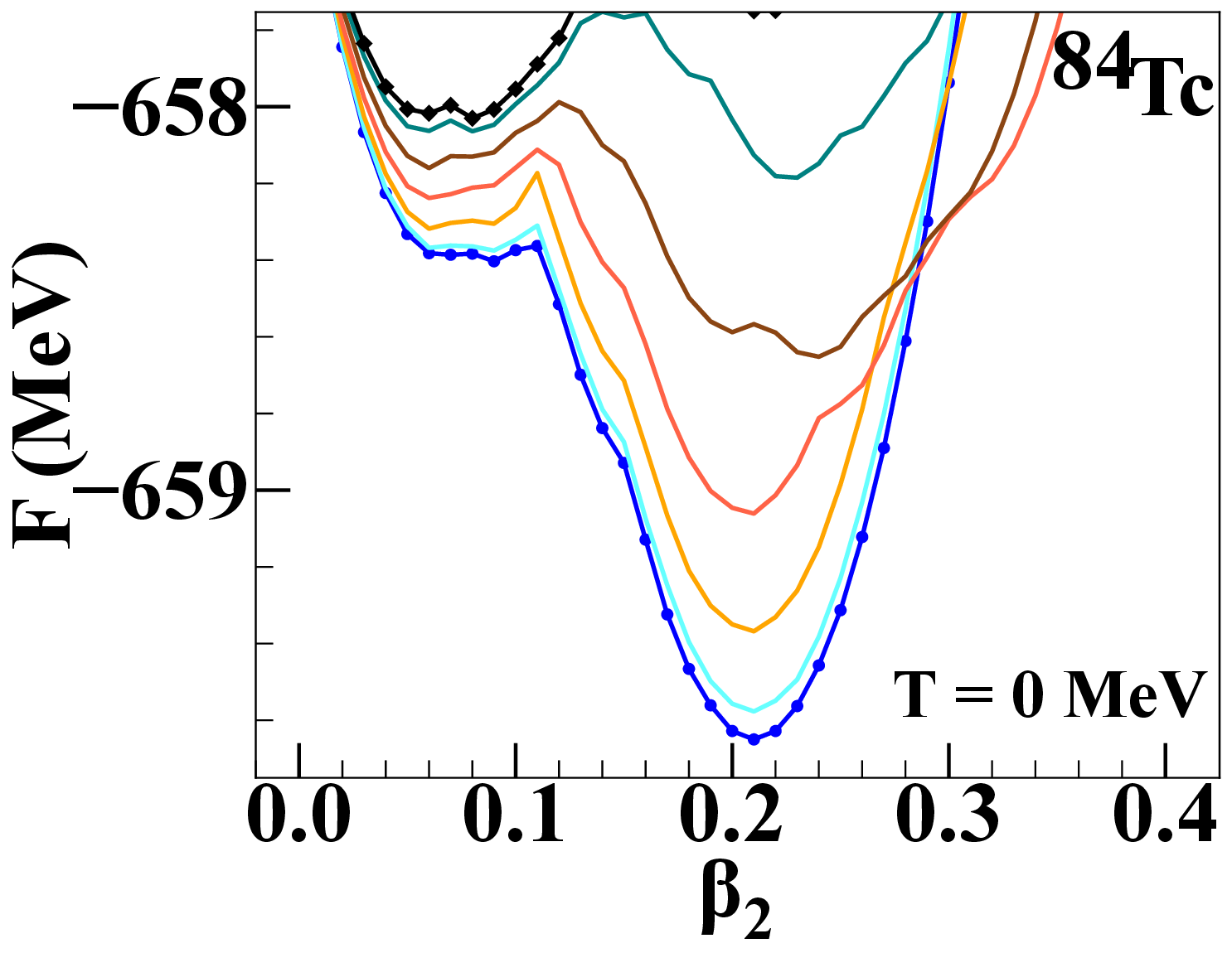, 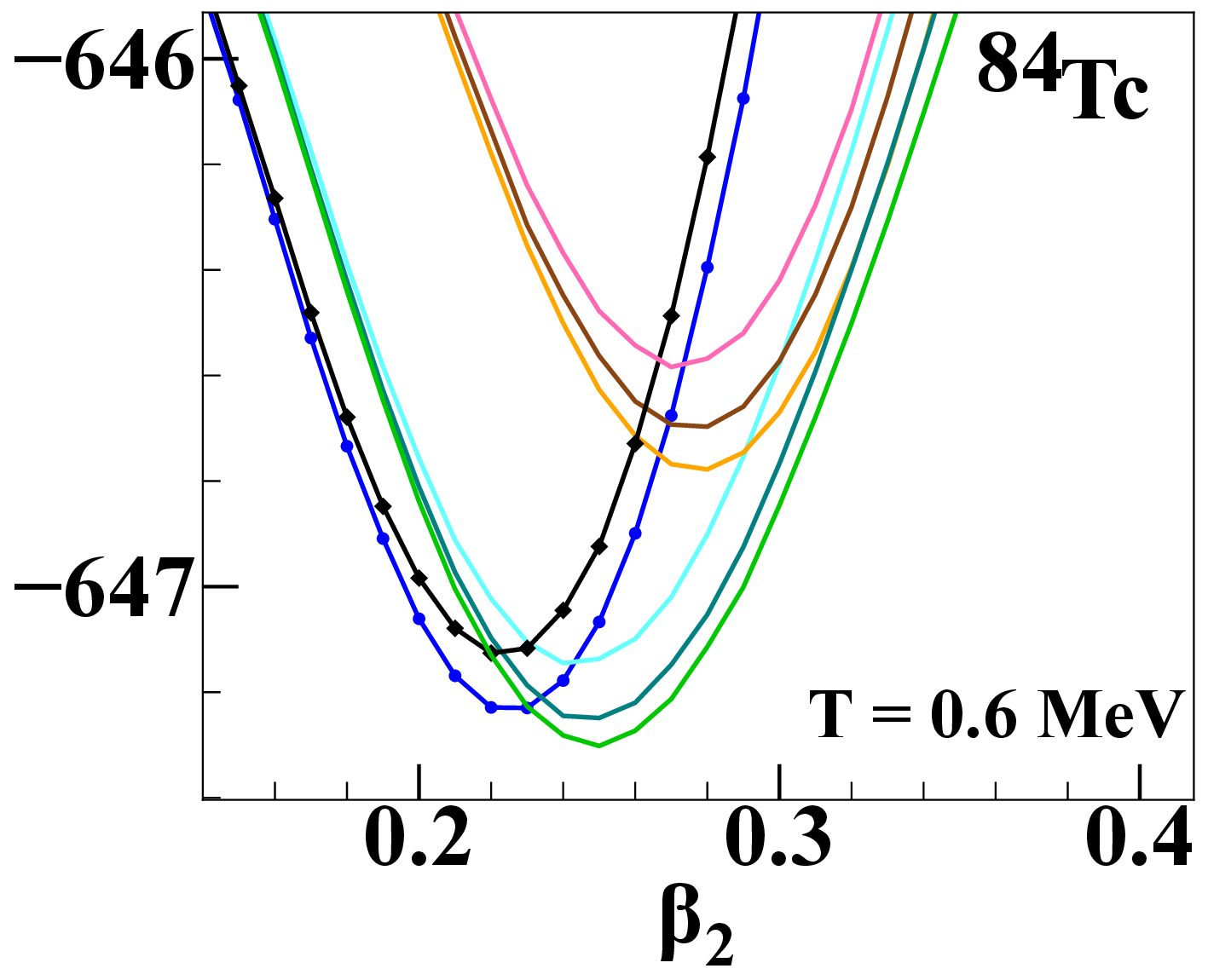, 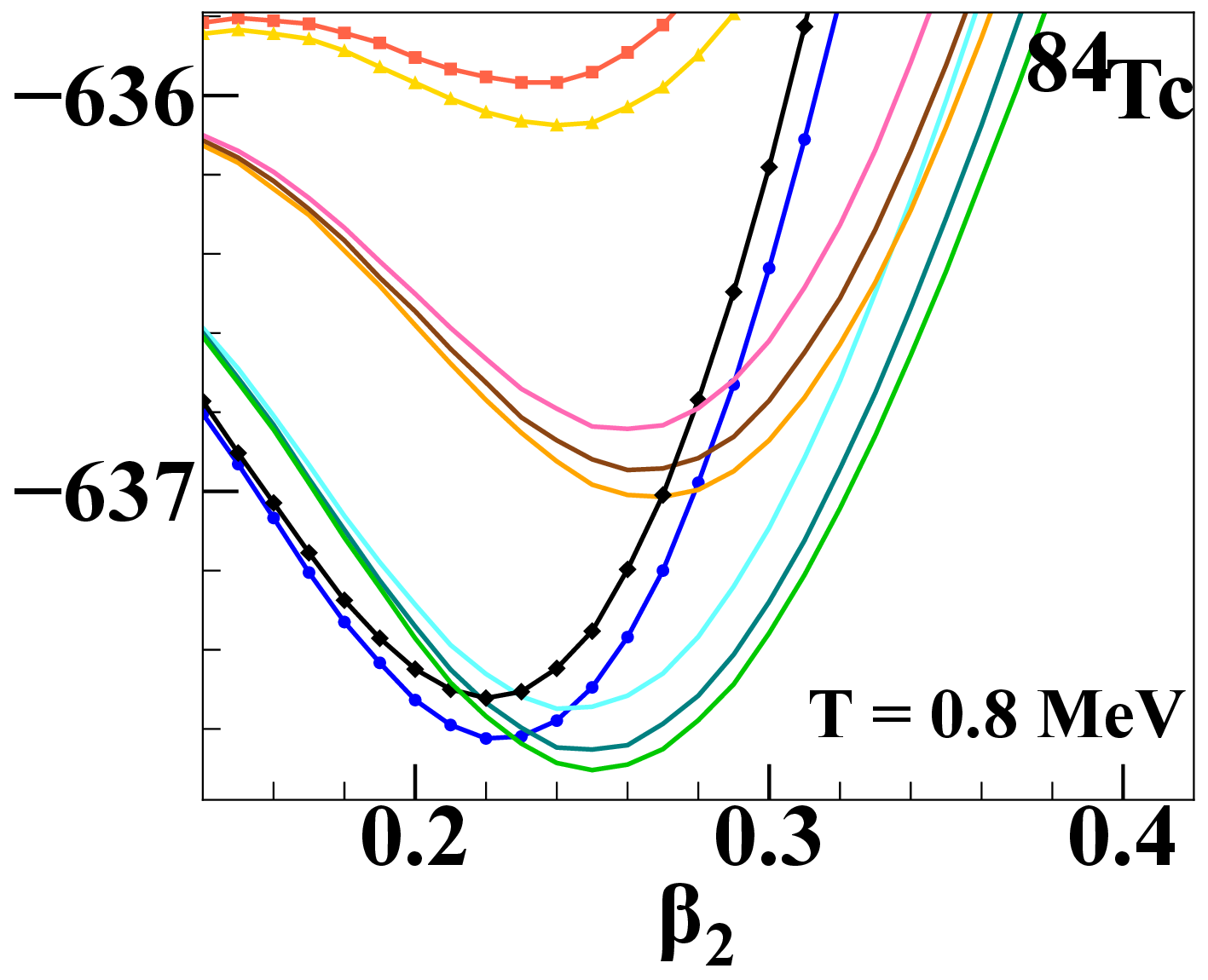,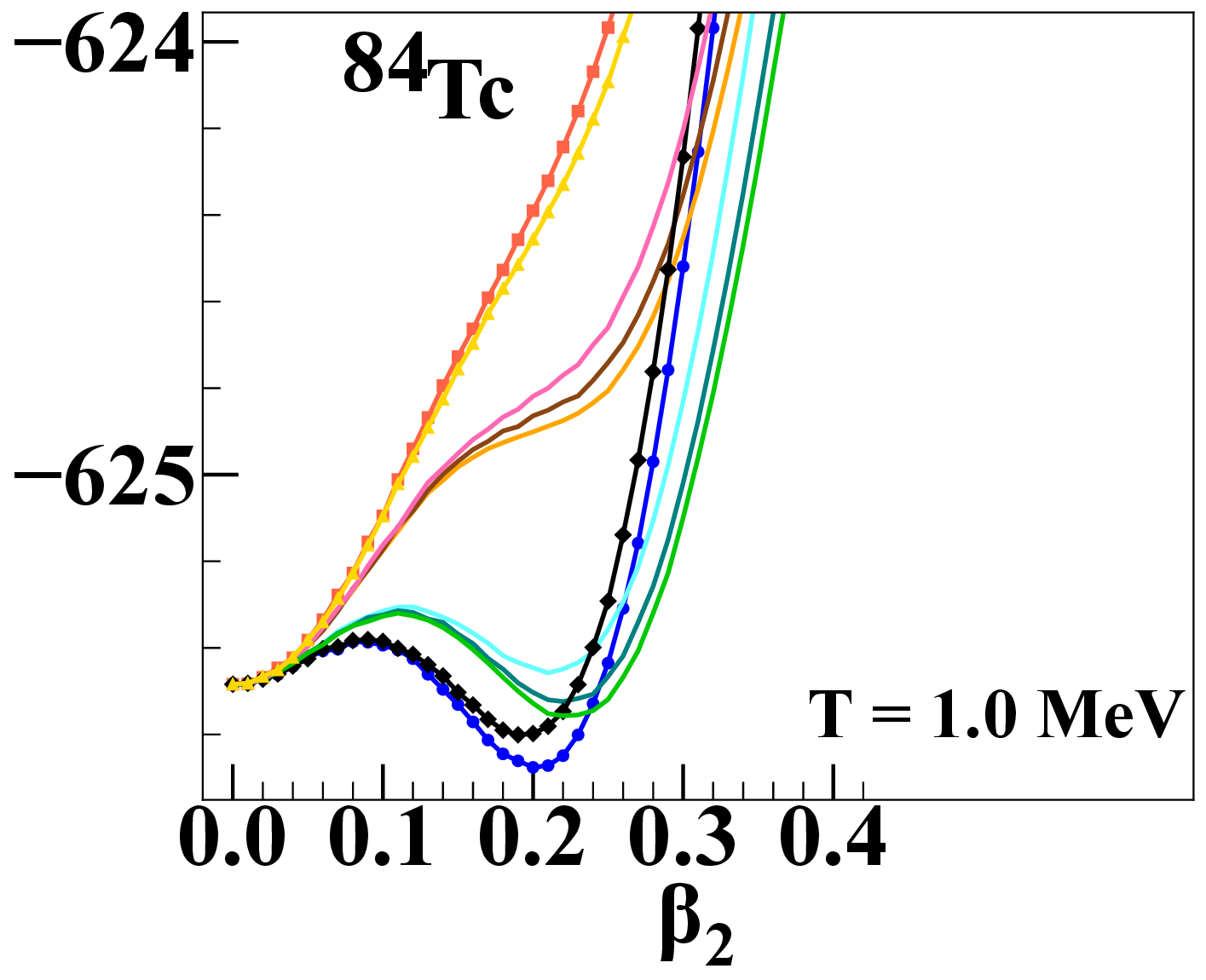}
     {
        \node[anchor=north west, inner sep=0] (s2\k) at ({\k*\w},-\h-0.2cm)
          {\includegraphics[width=\w,height=\h]{\c}};
      }

    % tall image at right (span both rows, automatic scale)
%    \node[anchor=north west, inner sep=0] (tall) at ({4*\w + 0.5cm},0)
%      {\includegraphics[height=\dimexpr 1\h-2cm, keepaspectratio]{example-image}};
\node[anchor=north east, inner sep=0] (tall) at ({3.53*\w + 1.5cm}, -0.7)
  {\includegraphics[height=4.cm,keepaspectratio]{legend.eps}};

  \end{tikzpicture}

  \caption{Shape evolution of $^{84}\mathrm{Ru}$ and its $\beta^+$decay daughter $^{84}\mathrm{Tc}$}
    \label{td84RuTc}
\end{figure}

The temperature effects on the shape evolution with axially symmetric and triaxial shapes is shown in the figures \ref{td108MoTc} and \ref{td84RuTc} (top row), where the free‑energy surfaces $F(\beta,\gamma)$ of $^{108}\mathrm{Mo}$ and $^{84}\mathrm{Ru}$ nuclei are presented for temperatures T $=$  0 to 1 MeV. These nuclei are found to exhibit $\beta^{-}$ and $\beta^{+}$ decay respectively by inter-changing one proton and neutron. Although $\gamma$ varies from $-180^\circ$ to $0^\circ$, only a few curves of $\gamma$ close to minima are presented in the free energy plots. The bottom row of figures \ref{td108MoTc} and \ref{td84RuTc} show shape evolution of their decay daughters $^{108}\mathrm{Tc}$, $^{84}\mathrm{Tc}$ for T = 0 to 1 MeV. Tracing of energy minima as a function of deformation parameters $\beta_2$ and $\gamma$ presented for determining shape and shape phase transitions in these hot isotopes show shape mixing and transitions with closely competing $\gamma$s. \par

The parent nucleus $^{108}\mathrm{Mo}$ is well deformed as seen in Fig.~\ref{td108MoTc}. The nearly oblate minimum ($\beta_2=0.24$, $\gamma=-170^\circ$) with a coexisting triaxial minimum ($\beta_2=0.3$, $\gamma=-140^\circ$) at T $=$ 0 MeV evolves with temperature, and transitions to a  triaxial shape at T $=$ 0.6 MeV. At T $=$ 0.8 MeV, the oblate and triaxial configurations with $\beta_2=0.27$ become nearly degenerate before the shape coexistence vanishes for T $\geq$ 1 MeV where single oblate minimum is seen. The initial rise in deformation with increasing T due to particle rearrangement near Fermi level is evident in Fig.~\ref{td108MoTc}.

The $\beta^-$ decay of $^{108}\mathrm{Mo}$ produces $^{108}\mathrm{Tc}$, which shows a triaxial minimum ($\gamma=-140^\circ$, $\beta_2=0.3$) coexisting with a nearly oblate ground state ($\gamma=-170^\circ$, $\beta_2=0.24$). With increasing temperature, it follows the same trend as $^{108}\mathrm{Mo}$, attaining a well-defined oblate shape ($\gamma=-180^\circ$) with decreasing deformation at T $=$ 1 MeV.

The shape evolution of $^{84}\mathrm{Ru}$ displayed in the first row of Fig.~\ref{td84RuTc} shows well deformed oblate minimum ($\beta_2=0.20$, $\gamma=-180^\circ$) with two shallow triaxial minima at T $=$ 0 MeV. At T $=$ 0.6 MeV, the lowest triaxial minimum ($\beta_2=0.25$, $\gamma=-40^\circ$) competes closely with the oblate configuration ($\beta_2=0.22$, $\gamma=-180^\circ$) and moves to deeper oblate shape ($\beta_2=0.22$) with shallower triaxial minimum ($\beta_2=0.24$, $\gamma=-40^\circ$) at T $=$ 0.8 MeV. For T $\geq$ 1 MeV, the coexistence vanishes as $^{84}\mathrm{Ru}$ attains a spherical configuration due to the washing out of shell effects.\par

The $\beta^+$ daughter $^{84}\mathrm{Tc}$, shown in the second row of Fig.~\ref{td84RuTc}, exhibits a shape evolution similar to its parent $^{84}\mathrm{Ru}$ at lower temperatures. However, at T $=$ 0.8 MeV the lowest minimum is triaxial ($\beta_2=0.25$, $\gamma=-40^\circ$), nearly degenerate with the oblate shape ($\beta_2=0.22$, $\gamma=-180^\circ$). At T $=$ 1.0 MeV, the nucleus stabilizes in an oblate configuration ($\beta_2=0.20$, $\gamma=-180^\circ$), indicates the shell-quenching effects will follow at higher temperatures.

 \begin{figure}[htbp]
 	\begin{subfigure}[t]{0.5\linewidth}
 		\centering
 		\includegraphics[width=\linewidth]{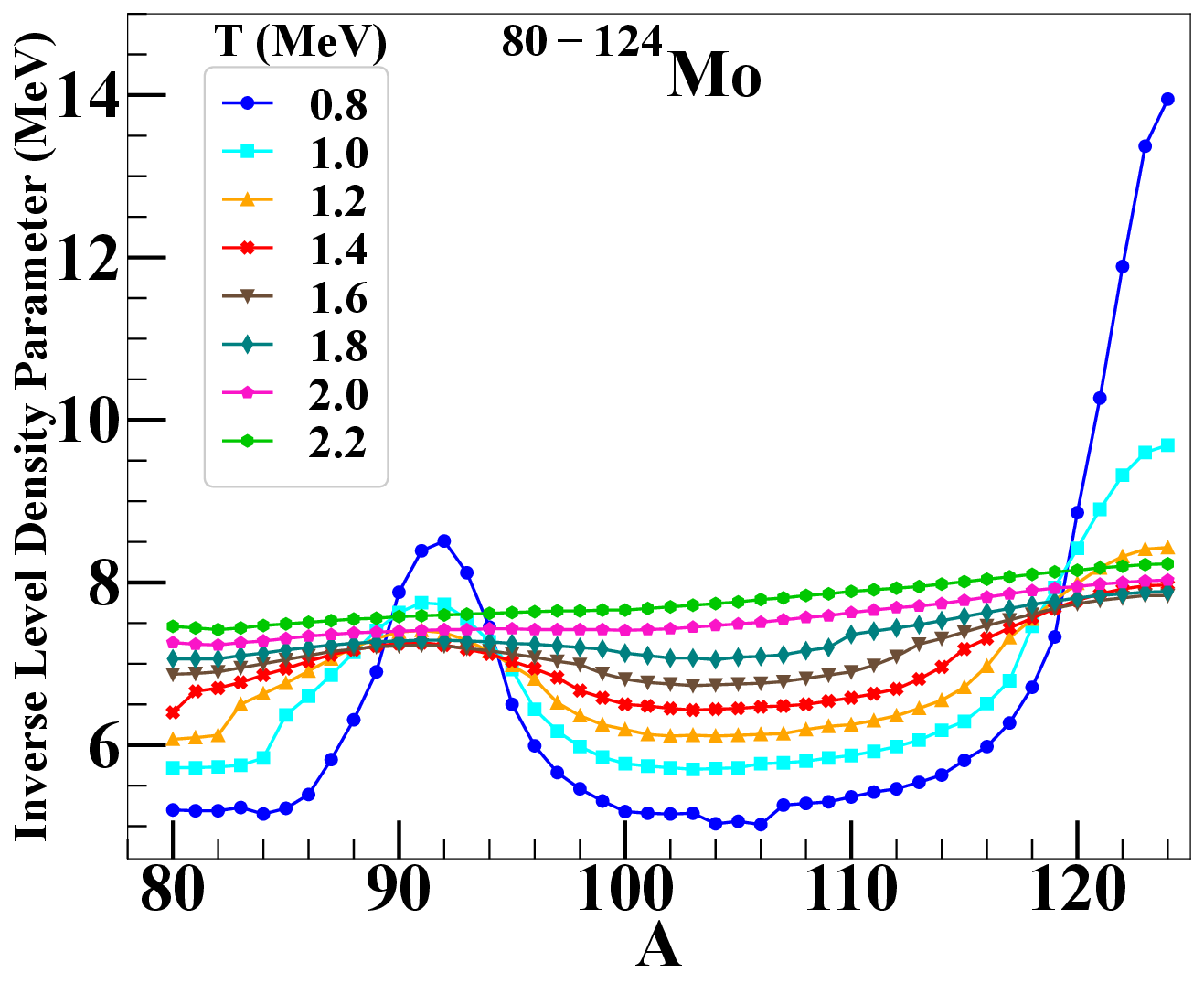}
 		 \caption{Mo Isotopes}
 		 \label{tkMo}
 	\end{subfigure}
 	\begin{subfigure}[t]{0.5\linewidth}
 		\centering
 		\includegraphics[width=\linewidth]{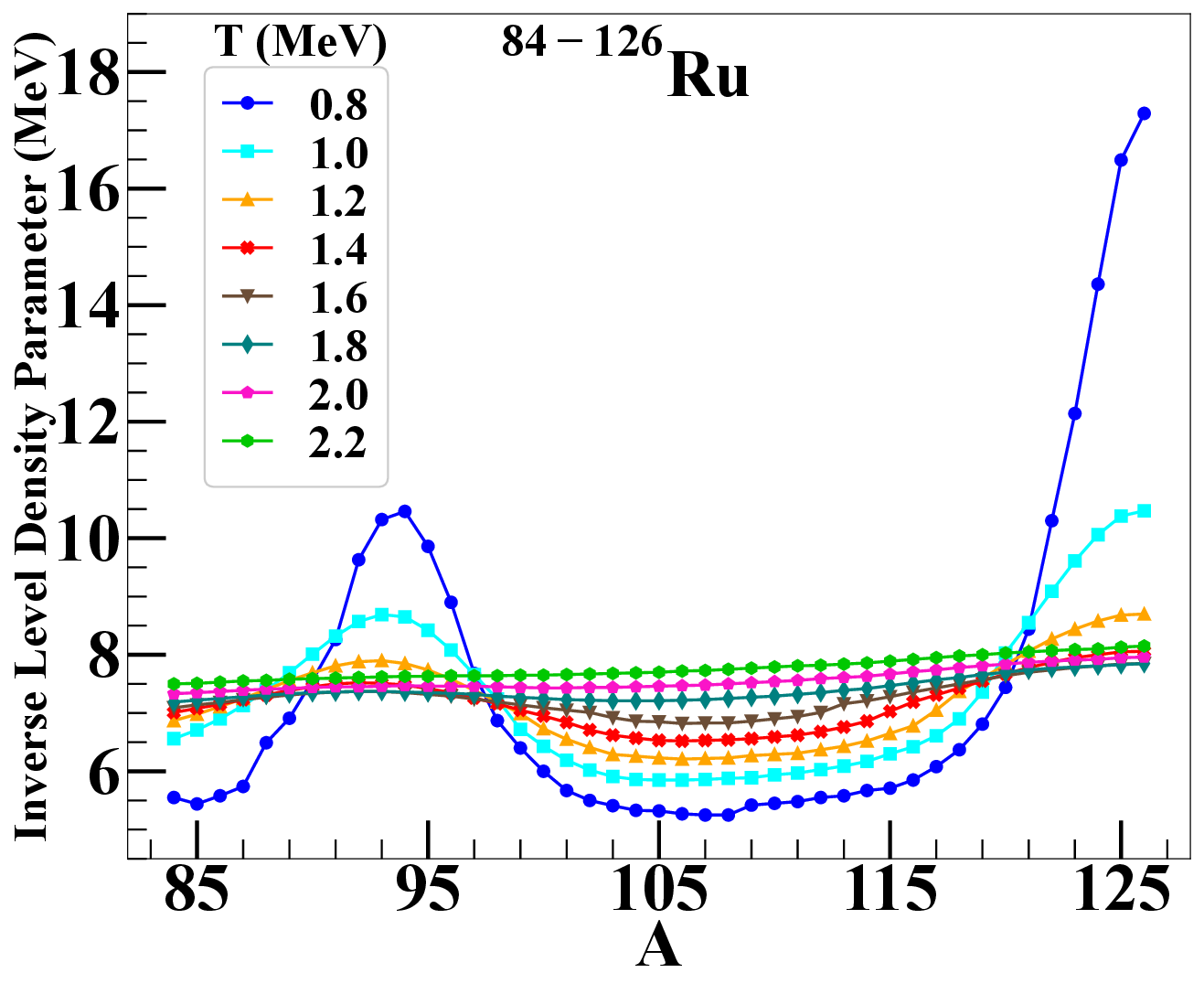}
 		\caption{Ru Isotopes}
 		 \label{tkRu}
 	\end{subfigure}
 	\caption{Variation of Inverse Level Density parameter (K) with Temperature.}
 	\label{Tk}
 \end{figure}

Level Density (LD) parameter 'a' is an important quantity directly related to the density of single-particle states near the Fermi surface which is strongly influenced by the shell effects at low energies. Fig.~\ref{Tk} displays the inverse level density parameter ($K=A/a$) for $^{80-124}\mathrm{Mo}$ and $^{84-126}\mathrm{Ru}$ isotopes at different temperatures between T $=$ 0.8 and 2.2 MeV. The level density parameter attains a minimum at shell closures, hence K shows  maxima near shell closures N =50, 82, as seen in Fig.~\ref{Tk}. The values of 'a' $\approx$ A/8 to A/11 are in good agreement with ref. \cite{ChankovaPRC73, RahmatinejadPRC101}. The continuum corrections become important only after T = 3 MeV as illustrated in \cite{Agrawal98} therefore not included here. Variation of K shows influence of shell effects at low T and quenching of shell effects with increasing T where K remains almost constant at around 8 MeV for all nuclei at high temperatures around 2 MeV. \par

\begin{table}[ht!]
\centering
\caption{Q-values for $\beta$-decay transitions.}
\label{QvalueTable}
\begin{tabular}{ccccc}
\toprule
\bf{T} & \bf{Nucleus} & \bf{Transition} & $Q_{\beta^-} $ & \bf{$Q_{\beta^-}$(Exp.)} \\
\bf{(MeV)} & & & \bf{(MeV)} & \bf{(MeV)}\\
\midrule
\multirow{4}{*}{0.8} & \multirow{4}{*}{$^{108}\mathrm{Mo}$} 
 & $G_p \to G_d$   & 5.2125 &  \multirow{4}{*}{5.174}\\
 & & $G_p \to S_d$  & 4.8825 &  \\
 & & $S_p \to G_d$      &5.6425        &  \\
 & & $S_p \to S_d$          & 5.3125 &  \\
\midrule
\multirow{4}{*}{0.6} & \multirow{4}{*}{$^{112}\mathrm{Mo}$} 
 & $G_p \to G_d$				& 7.8624 & \multirow{4}{*}{7.78}  \\
 & & $G_p \to S_d$              & 7.6524 &  \\
 & & $S_p \to G_d$              & 8.1324 &  \\
 & & $S_p \to S_d$           & 7.9224 & \\
 \midrule
\multirow{4}{*}{0.8} & \multirow{4}{*}{$^{112}\mathrm{Ru}$} 
  & $G_p \to G_d$ & 4.2125 & \multirow{4}{*}{4.1}  \\
 & & $G_p \to S_d$             & 4.0025 &  \\
 & & $S_p \to G_d$              & 4.4125 &  \\
 & & $S_p \to S_d$           & 4.2025 &  \\
\midrule
\midrule
\bf{T} & \bf{Nucleus} & \bf{Transition} & \bf{$Q_{\beta^+}$} & \bf{$Q_{\beta^+}$(Exp.)} \\
\bf{(MeV)} & & & \bf{(MeV)} & \bf{(MeV)}\\
\midrule
\multirow{4}{*}{0.6} & \multirow{4}{*}{$^{84}\mathrm{Ru}$} 
& $G_p \to G_d$				& 12.5855 &  \multirow{4}{*}{Not Available} \\
 & & $G_p \to S_d$              & 12.7955 & \\
 & & $S_p \to G_d$             & 12.4155 &  \\
 & & $S_p \to S_d$           & 12.6255 & \\
\bottomrule
\end{tabular}
\end{table}

Beta-decay, a weak-interaction process, is used as a probe of nuclear structure and as a key input in nucleosynthesis r-process. In our earlier study \cite{MANPA24} we have demonstrated how shape-coexisting parent and daughter nuclei influence decay Q-values which affect life-times. Here, we include finite temperature to our analysis of decay Q-values for studying changes due to competing minima of different shapes. \par 

The potential-energy surface for nuclei $^{108}\mathrm{Mo}$, $^{112}\mathrm{Mo}$, $^{112}\mathrm{Ru}$ and $^{84}\mathrm{Ru}$ (shown in Figs \ref{td108MoTc}, \ref{td84RuTc}) at temperatures T $=$ 0.6 and 0.8 MeV, exhibit two competing minima of different shapes: the lowest-energy minimum (denoted by G) and a second minimum (S) with slightly higher energy. This coexistence gives rise to four possible decay paths, depending on whether the parent and daughter nuclei occupy their first (G) or second (S) minimum: first minimum of parent (denoted by $G_p$) to first minimum of daughter ($G_d$): ($G_p \to G_d$); first minimum of parent ($G_p$) to second minimum of daughter ($S_d$): ($G_p \to S_d$); second minimum of parent ($S_p$) to first minimum of daughter ($G_d$): ($S_p \to G_d$); and second minimum of parent ($S_p$) to second minimum of daughter ($S_d$): ($S_p \to S_d$). Q-values for $\beta$-decay through these four possible decay paths are calculated in Table~\ref{QvalueTable}. As binding energies for these minima differ by a few hundred KeV, we observe variations in Q-values, depending on which minima are involved in the decay. The calculated $Q$-values for $\beta$-decay transitions are compared with the experimental values obtained from the 'QCalc tool' at NNDC \cite{nndc}, which are based on mass values from on the 2020 Atomic Mass Evaluation (AME2020). As seen in Table~\ref{QvalueTable}, the theoretical values agree reasonably with the experimental $Q_{\beta^-}$ values.  Experimental $Q_{\beta^+}$ for $^{84}\mathrm{Ru}$ is not available for comparison, as this nucleus lies close to the proton drip line.

The $Q_{\beta^-}$ value of $G_p \to G_d$ transition for $^{108}\mathrm{Mo}$ and  $^{112}\mathrm{Mo}$, $Q_{\beta^-}$ of  the $S_p \to G_d$ transition for $^{112}\mathrm{Ru}$, agree most closely with the experimental Q-value. We observe, the $G_p \to G_d$ transition is not always the closest to experimental values and other transitions, such as $S_p \to G_d$, can better reproduce the experimental $Q_{\beta}$ values, indicating that observed transitions might take place from these minimas.

These results illustrate that shape coexistence influences decay dynamics at finite temperature. Q-values vary depending on whether the parent and daughter occupy the same or different shape minima. The interplay between $Q_{\beta^{\pm}}$ value, and the transition mode emphasizes the influence of shape evolution and shape coexistence on decay at finite temperatures. This highlights the necessity of incorporating the temperature effects into decay models, especially when competing shapes lie close in energy for which a reliable theoretical model is essential. This is a preliminary study on temperature effects on nuclear stability. More detailed calculations and results on lifetime and shape coexistence will be presented in our upcoming communication.  \par

\section{Conclusion}
To conclude, the temperature effects in $^{80-124}\mathrm{Mo}$, and $^{84-126}\mathrm{Ru}$ isotopes have been studied in a theoretical framework using  macroscopic-microscopic approach and statistical theory of hot nuclei. This region known for rapid shape transitions and shape mixing shows various shape coexisting states with mostly oblate and triaxial shapes. At low $T$, the near degeneracy oblate and triaxial configurations in $^{108}\mathrm{Mo}$, $^{84}\mathrm{Ru}$, $^{108}\mathrm{Tc}$, and $^{84}\mathrm{Tc}$ indicates low‑lying shape‑coexisting bands and enhanced shape mixing. Thermal effects wash away shape mixing and quench shape coexistence as shown in this work. The second minimum state in shape coexisting states, influences structure, decay modes and beta-decay Q-value.\par

\section*{Acknowledgement}
Mamta Aggarwal and G. Saxena are grateful for the support 
received from DST, Government of India, under DST/WIDUSHI-B/PM/2024/23, and the Department of Science \& Technology (DST), Government of Rajasthan, Jaipur, India under F24(1)DST/RandD/2024-EAC-00378-6549873/819 respectively.

\end{document}